\documentclass[5p,twocolumn]{elsarticle}
\usepackage{amssymb}
\usepackage{graphicx}
\usepackage{hyperref}
\usepackage{verbatim}

\journal{Computer Physics Communications}

\begin{document}

\title{QuTiP: An open-source Python framework for the dynamics of open quantum systems}

\author[riken]{J. R. Johansson\fnref{fn1}\corref{cor1}}
\ead{robert@riken.jp}
\author[riken,um]{P. D. Nation\fnref{fn1}\corref{cor1}}
\ead{pnation@riken.jp}
\author[riken,um]{Franco Nori}
\address[riken]{Advanced Science Institute, RIKEN, Wako-shi, Saitama 351-0198, Japan}
\address[um]{Department of Physics, University of Michigan, Ann Arbor, Michigan 48109-1040, USA}
\fntext[fn1]{These authors contributed equally to this work.}
\cortext[cor1]{Corresponding authors}
\date{\today}

\begin{abstract}
We present an object-oriented open-source framework for solving the dynamics of open quantum systems written in Python.  Arbitrary Hamiltonians, including time-dependent systems, may be built up from operators and states defined by a quantum object class, and then passed on to a choice of master equation or Monte-Carlo solvers.  We give an overview of the basic structure for the framework before detailing the numerical simulation of open system dynamics.  Several examples are given to illustrate the build up to a complete calculation.  Finally, we measure the performance of our library against that of current implementations.  The framework described here is particularly well-suited to the fields of quantum optics, superconducting circuit devices, nanomechanics, and trapped ions, while also being ideal for use in classroom instruction.
\end{abstract}

\maketitle
{\bf PROGRAM SUMMARY}
\begin{small}
\noindent
\\
{\em Manuscript Title:} QuTiP: An open-source Python framework for the dynamics of open quantum systems  \\
{\em Authors:} J. R. Johansson, P. D. Nation \\
{\em Program Title:} QuTiP: The Quantum Toolbox in Python   \\
{\em Journal Reference:}  \\
  %Leave blank, supplied by Elsevier.
{\em Catalogue identifier:}   \\
  %Leave blank, supplied by Elsevier.
{\em Licensing provisions:} GPLv3 \\
  %enter "none" if CPC non-profit use license is sufficient.
{\em Programming language:} Python   \\
{\em Computer:} i386, x86-64  \\
  %Computer(s) for which program has been designed.
{\em Operating system:} Linux, Mac OSX, Windows  \\
  %Operating system(s) for which program has been designed.
{\em RAM:} 2+~Gigabytes                                              \\
  %RAM in bytes required to execute program with typical data.
{\em Number of processors used:} 1+                              \\
  %If more than one processor.
  % Please give some freely chosen keywords that we can use in a
  % cumulative keyword index.
{\em Classification:} 7 Condensed Matter and Surface Science   \\
{\em External routines/libraries:} NumPy, SciPy, Matplotlib \\
  % Fill in if necessary, otherwise leave out.
{\em Nature of problem:} Dynamics of open quantum systems \\
  %Describe the nature of the problem here.
{\em Solution method:} Numerical solutions to Lindblad master equation or Monte-Carlo wave function method.\\
  %Describe the method solution here.
{\em Restrictions:} Problems must meet the criteria for using the master equation in Lindblad form.\\
  %Describe any restrictions on the complexity of the problem here.
{\em Running time:} A few seconds up to several tens of minutes, depending on size of underlying Hilbert space.\\
  %Give an indication of the typical running time here.
   \\
\end{small}
\section{Introduction}
Every quantum system encountered in the real world is an open quantum system \cite{breuer:2002}.  For although much care is taken experimentally to eliminate the unwanted influence of external interactions, there remains, if ever so slight, a coupling between the system of interest and the external world.  In addition, any measurement performed on the system necessarily involves coupling to the measuring device, therefore introducing an additional source of external influence.  Consequently, developing the necessary tools, both theoretical and numerical, to account for the interactions between a system and its environment is an essential step in understanding the dynamics of quantum systems.

By definition, an open quantum system is coupled to an environment, also called a reservoir or bath, where the complexity of the environmental dynamics renders the combined evolution of system plus reservoir intractable.  However, for a system weakly coupled to its surroundings, there is a clear distinction between the system and its environment, allowing for the dynamics of the environment to be traced over, resulting in a reduced density matrix describing the system alone.  The most general dynamical equation governing this reduced system density matrix is given by the Lindblad master equation \cite{lindblad:1976} describing the evolution of an ensemble average of a large (formally infinite) number of identical system realizations.  Although the density operator formalism sufficed for the first half-century of quantum mechanics, the advent of single-ion traps in the 1980's \cite{horvath:1997} motivated the study of the quantum trajectories, or Monte-Carlo, description for the evolution of a \textit{single} realization of a dissipative quantum system \cite{plenio:1998}.

In general, for all but the most basic of Hamiltonians, an analytical description of the system dynamics is not possible, and one must resort to numerical simulations of the equations of motion.  In absence of a quantum computer \cite{buluta:2011}, these simulations must be carried out using classical computing techniques, where the exponentially increasing dimensionality of the underlying Hilbert space severely limits the size of system that can be efficiently simulated \cite{feynman:1982, buluta:2009}.  However, in many fields such as quantum optics \cite{haroche:2006,obrien:2009}, trapped ions \cite{horvath:1997,blatt:2008}, superconducting circuit devices \cite{you:2005,schoelkopf:2008,you:2011}, and most recently nanomechanical systems \cite{armour:2008,blencowe:2008,oconnell:2010,teufel:2011}, it is possible to design systems using a small number of effective oscillator and spin components, excited by a small number of quanta, that are amenable to classical simulation in a truncated Hilbert space.

Of the freely available quantum evolution software packages \cite{schack:1997,tan:1999,vukics:2007}, the Quantum Optics Toolbox (qotoolbox) \cite{tan:1999} has by far been the most successful.  Although originally geared toward quantum optics, the qotoolbox has gained popularity in a variety of disciplines, driven in part by its rapid modeling capabilities and easy to read code syntax.  Yet, at the same time, the qotoolbox has not been updated in nearly a decade, leaving researchers to rely on an outdated numerical platform.  Moreover, while the code underlying the qotoolbox is open-sourced, it does rely on the proprietary Matlab \cite{matlab} computing environment making it an impractical solution for many research groups, as well as for use as an educational tool inside the classroom.

In this paper, we describe a fully open-source implementation of a framework designed for simulating open quantum dynamics written in the Python programming language \cite{python} called the \textit{Quantum Toolbox in Python} or QuTiP \cite{qutip}.  This framework distinguishes itself from the other available software solutions by providing the following advantages:

\begin{itemize}
\item Based entirely on open-source software.

\item Easy to read, rapid code development using the Python programming language.

\item Support for arbitrary, time dependent Hamiltonians.

\item Makes use of the multiple processing cores found in modern computers.

\item Community based infrastructure, allowing for user contributions to the code base. 
\end{itemize}

Although Python is an interpreted programming language, it is well suited for scientific computations as a result of its large collection of high-performance low-level numerical libraries \cite{oliphant:2007}, mathematical functions \cite{scipy}, and data visualization capabilities \cite{matplotlib}, that largely are implemented in efficient compiled code. In particular, QuTiP relies heavily on the sparse matrix and dense array functionality provided by the {\it SciPy} \cite{scipy} and {\it NumPy} \cite{oliphant:2007} packages, respectively.  Since the bulk of a typical calculation is spent in these libraries, a QuTiP simulation can achieve nearly the same performance as compiled code. The advantage of using the Python programming language over a compiled programming language is a greatly simplified development process, and more transparent, less bug-prone code. For data visualization QuTiP uses the matplotlib package \cite{matplotlib}, which is capable of producing publication-quality 2D and 3D figures in a wide range of styles and formats.

Given the success of the qotoolbox, the development of QuTiP has in part been directed toward providing a replacement for this excellent, yet aging software. In the spirit of open-source development, we have strived to use the best parts of the qotoolbox in QuTiP, while improving, replacing, or complementing the parts that were in need of modernization. The result is a framework for simulating quantum system dynamics that is in many ways more efficient and better suited for modern computers, as well as better positioned for further development and adoption to new computer architecture advances.  Given the size of the QuTiP framework, we do not hope to cover all of its functionality here.  Instead, we will focus on the key data structures, and numerical routines underlying the majority of calculations. In addition, we will highlight a variety of example calculations that we hope will give the reader a flavor of the capabilities of QuTiP, and highlight what is possible using this framework.  A complete overview of QuTiP is given on its website \cite{qutip}.

This paper is organized as follows.  In Sec.~\ref{sec:framework} we introduce the main QuTiP class, representing a quantum operator or state vector, and its associated data structures and methods.  In Sec.~\ref{sec:evolution} we give a brief overview of the density matrix formalism before discussing the master equation and Monte-Carlo methods used in QuTiP.  Section \ref{sec:numerical} presents a selection of examples meant to illustrate how calculations are performed using the QuTiP framework.  Section~\ref{sec:performance} compares the performance of the QuTiP master equation and Monte-Carlo solvers to those in the qotoolbox.  Finally, Sec.~\ref{sec:conclusion} briefly concludes, while a list of user accessible functions built into QuTiP, as well as example codes, are relegated to the appendix.

\section{The QuTiP framework}\label{sec:framework}

QuTiP provides an object-oriented framework for representing generic quantum systems, and for performing calculations and simulations on such systems.  In order to simulate a quantum system, one must first construct an object that encapsulates the properties of an arbitrary state vector or operator.   
\begin{figure}[t]
\begin{center}
\includegraphics[width=8.5cm]{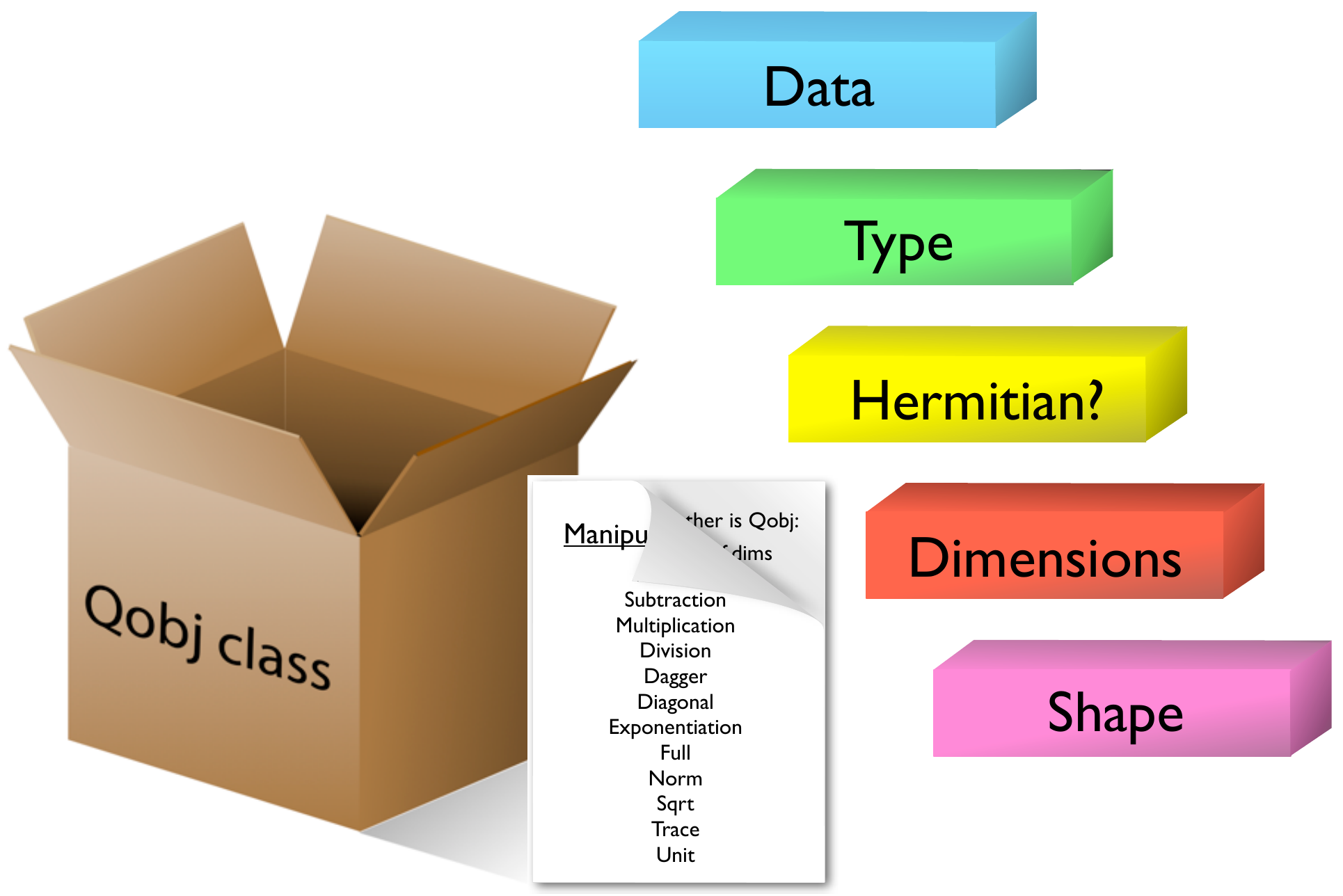}
\caption{(Color) A quantum object, operator or state vector, is represented by a quantum object class (\texttt{Qobj}) instance.  The \texttt{Qobj} class may be thought of as a container, holding the data structures required to fully characterize a generic quantum object, as well as a list of instructions on how to manipulate these items. The primary data structures are the data, in sparse matrix form, that represents a quantum object in a given Hilbert space, the type of the object (ket, bra, operator, super-operator), whether it is Hermitian or not, and the objects dimension and shape. Here, the dimension describes the structure of the Hilbert space, i.e., whether it is a composite system and how it is composed. The \texttt{Qobj} class also defines a variety of methods that implement common functions operating on quantum objects. See Table~\ref{tbl:methods} for a list of the \texttt{Qobj} class methods.}
\label{fig:qobj}
\end{center}
\end{figure}
A unified representation of quantum operators and state vectors is implemented in QuTiP by means of the quantum object class (\texttt{Qobj}), that uses a sparse matrix representation of a quantum object in a finite dimensional Hilbert space. The \texttt{Qobj} class internally maintains a record of the principal attributes of the quantum object it represents.  These include, the objects type (i.e ket, bra, operator, or super-operator), whether the underlying object is Hermitian, the dimensionality of a composite object formed via the tensor-product, and the size of the sparse data matrix.  A schematic illustration of the key components underlying the \texttt{Qobj} class is shown in Fig.~\ref{fig:qobj}.

In addition to serving as a book-keeper for the properties of a quantum object, the \texttt{Qobj} class is also a computational object, implementing the usual binary arithmetic operations, and a variety of class methods for performing common object manipulations as presented in Table~\ref{tbl:methods}.
\begin{table}[b]
\begin{center}
\begin{tabular}{ll}
Method & Description
\\ \hline\hline
\texttt{dag()} & Adjoint of the quantum object.
\\ \hline
\texttt{diag()} & Diagonal elements of object.
\\ \hline
\texttt{eigenstates()} & Eigenstates and eigenvectors.
\\ \hline
\texttt{expm()} & Exponentiated quantum object.
\\ \hline
\texttt{full()} & Dense array representation.
\\ \hline
\texttt{norm()} & L2 norm (states), trace norm (oper).
\\ \hline
\texttt{sqrtm()} & Matrix square root.
\\ \hline
\texttt{tr()} & Trace of quantum object.
\\ \hline
\texttt{unit()} & Normalizes the quantum object.
\\ \hline
\end{tabular}
\end{center}
\caption{List of methods built into the \texttt{Qobj} class.}
\label{tbl:methods}
\end{table}
Therefore, with just a few lines of QuTiP code, it is straightforward to construct Hamiltonians from arbitrary combinations of operators, and to construct density matrices and state vectors that represent complicated superpositions of basis states. To further simplify this important step, QuTiP provides a library of commonly occurring operators and states which are given in \ref{sec:list}.  

For example, to create an instance of the \texttt{Qobj} class that represents the ubiquitous two-level Hamiltonian $(\hbar=1)$
\begin{eqnarray}\label{eq:twolevel}
   H = \frac{1}{2}\epsilon\sigma_z + \frac{1}{2}\Delta\sigma_x,
\end{eqnarray}
with energy splitting $\epsilon$ and transition energy $\Delta$, one can use the following QuTiP code:
\begin{footnotesize}
\begin{verbatim}
H = 0.5 * epsilon * sigmaz() + 0.5 * delta * sigmax()
\end{verbatim}
\end{footnotesize}
where \texttt{epsilon} and \texttt{delta} represent user defined constants.  The result is a single \texttt{Qobj} instance \texttt{H} that represents the Hamiltonian operator Eq.~(\ref{eq:twolevel}).

Composite quantum systems are nearly as easy to create. Consider the Jaynes-Cumming Hamiltonian
\begin{eqnarray}
\label{eq:hamiltonian_jc}
   H = \omega_0 a^\dag a  + \frac{1}{2}\epsilon\sigma_z + g (a^\dag \sigma_{+} +a \sigma_{-}),
\end{eqnarray}
with cavity frequency $\omega_{0}$ and coupling strength $g$, describing a cavity field coupled to a two-level atom (qubit).  A \texttt{Qobj} instance representing this composite Hamiltonian can be created with the following code:
\begin{footnotesize}
\begin{verbatim}
a  = tensor(destroy(N), qeye(2))
sm = tensor(qeye(N), destroy(2))
sz = tensor(qeye(N), sigmaz())
H  = omega0 * a.dag() * a + 0.5 * epsilon * sz 
     + g * (a.dag() * sm + a * sm.dag())
\end{verbatim}
\end{footnotesize}
where the \texttt{tensor} function is used to construct composite operators for the combined Hilbert space of the cavity (truncated to the $N$ lowest Fock states) and the atom. 

Since the \texttt{Qobj} class provides a unified representation for operators and states, we can use exactly the same technique to generate quantum states (either as state vectors or density matrices). A possible initial state for the system described by Eq.~(\ref{eq:hamiltonian_jc}), is generated with the QuTiP code:
\begin{footnotesize}
\begin{verbatim}
psi0 = tensor(fock(N,0),(fock(2,0)+fock(2,1)).unit())
\end{verbatim}
\end{footnotesize}
creating the \texttt{Qobj} representation of a cavity in its ground state, coupled to a qubit in a balanced superposition of its ground and excited state $\left|\Psi(0)\right>=(\left|0\right>_{c}\left|0\right>_{q}+\left|0\right>_{c}\left|1\right>_{q})/\sqrt{2}$.  Here, the subscripts $c$ and $q$ denote the cavity and qubit states, respectively.  The normalization factor ($\sqrt{2}$) is applied automatically using the \texttt{unit()} method.

In the previous example, we have used builtin QuTiP library functions to generate \texttt{Qobj} instances of commonly occurring operators and states, and the associated arithmetic operations in the \texttt{Qobj} class\footnote{The binary arithmetic operators +, -, and * are defined for two \texttt{Qobj} objects, and +, -, * as well as / are defined between a \texttt{Qobj} object and a real or complex number.} to combine these quantum operators into more complicated systems. The close correspondence between the mathematical formulation and the programming code makes it easy and transparent to define quantum objects. This is especially important when working with quantum systems of more complex structure.  Note that the \texttt{Qobj} instances in the previous examples are all self-contained descriptions of the quantum object they represent.  From the \texttt{Qobj} instance alone, a number of properties pertaining to the quantum object may be calculated, and various operations be applied (see Table~\ref{tbl:methods}). This includes, for example, operations such as the Hermitian adjoint, normalization, and trace, as well as  computation of the eigenstates and eigenenergies and operator exponentiation.

In addition, QuTiP includes several important functions operating on multiple states and/or operators (see Table~\ref{tbl:list}). We have already seen one such example in the \texttt{tensor} function used to generate tensor product states.  These states may also be decomposed into their  constituent parts by performing a partial trace over selected degrees of freedom using the function \texttt{ptrace}. For example, from the composite wave function \texttt{psi0} for the oscillator-qubit system, the state of the qubit can be extracted by tracing out the oscillator degrees of freedom using the QuTiP code:
\begin{footnotesize}
\begin{verbatim}
rho0_qubit = ptrace(psi0, 1)
\end{verbatim}
\end{footnotesize}
where the second argument is the index of the system that we wish to {\it keep}. In general, it can be a list of indices.  The properties of the resulting \texttt{Qobj} instance (shown in Fig.~\ref{fig:qobj}) may be inspected by displaying the string representation of the object returned by its \texttt{str} method. This method is implicitly invoked when the object is printed to the standard output 
\begin{footnotesize}
\begin{verbatim}
print rho0_qubit
Quantum object: dims = [[2], [2]], shape = [2, 2], 
type = oper, isHerm = True
Qobj data = 
[[ 0.5  0.5]
 [ 0.5  0.5]]
\end{verbatim}
\end{footnotesize}
which, in this case, shows that the \texttt{Qobj} instance \texttt{rho0\_qubit} is a $2\times2$, Hermitian quantum operator representing a balanced coherent superposition of its two basis states.  From a {\texttt Qobj} instance one may also calculate items such as the expectation value (\texttt{expect}) for an arbitrary operator with the QuTiP function, find the fidelity (\texttt{fidelity}) between two density matrices \cite{nielsen:2000}, or calculate the Wigner function (\texttt{wigner}) of a quantum state.  Using these, and other functions (Table.~\ref{tbl:list}), in the exploration of open quantum dynamics will be the focus of Sec.~\ref{sec:numerical}.

\begin{figure}[t]
\begin{center}
\includegraphics[width=8.3cm]{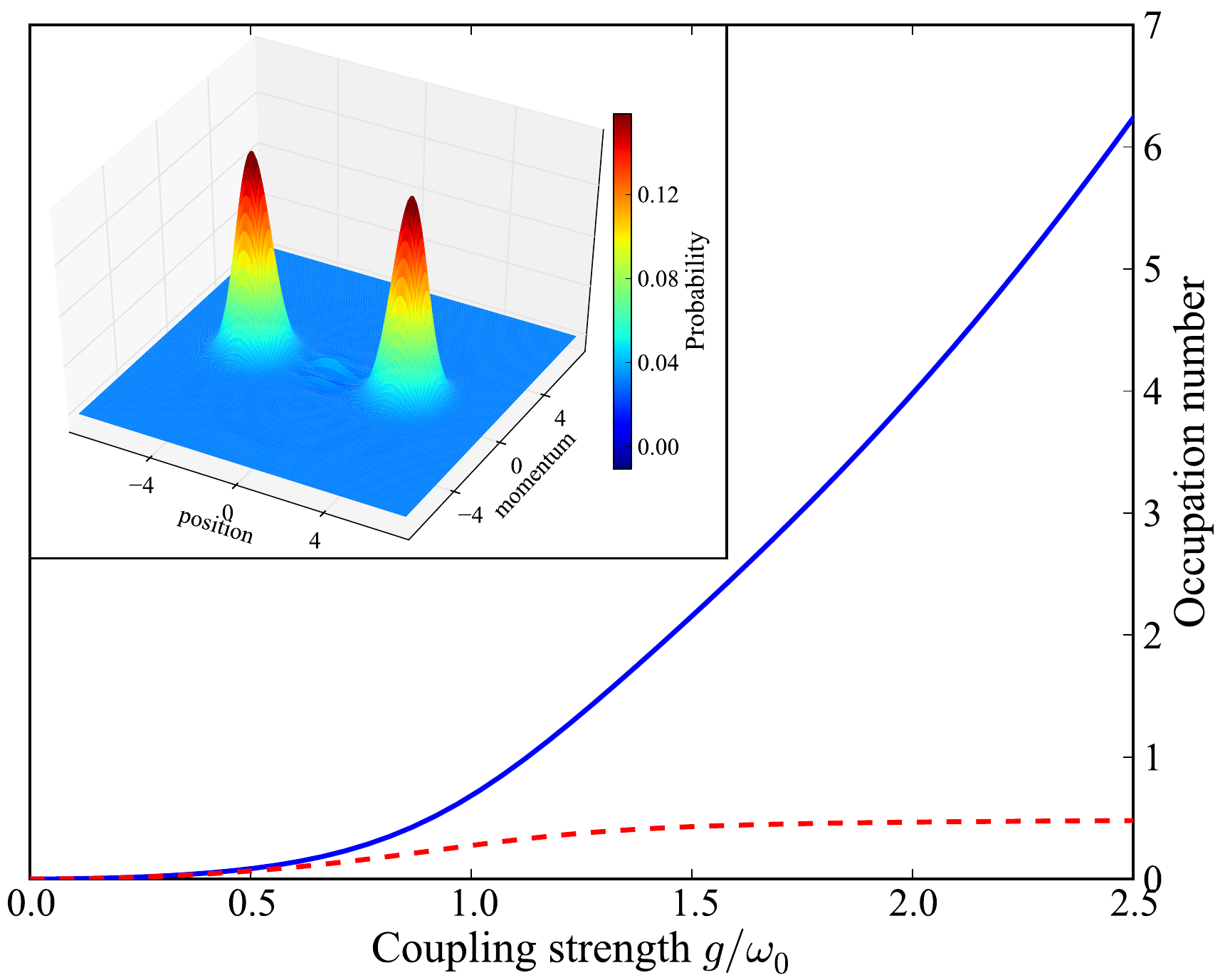}
\caption{(Color) Expectation value for the number of excitations in the cavity (blue) and qubit (dashed-red) modes of the non-RWA Jaynes-Cummings model Eq.~(\ref{eq:nonrwa}) as the coupling strength $g$ is increased into the ultra-strong coupling regime $g/\omega_{0}\gtrsim 1$.  The inset figure displays the Wigner function for the cavity mode at the largest coupling strength, $g=2.5\omega_0$, where $\omega_0$ is the bare cavity frequency.  At this coupling value, the state of the system is well-approximated by Eq.~(\ref{eq:approx}).}
\label{fig:nonrwa}
\end{center}
\end{figure}

Even though the emphasis of QuTiP is on dynamical modeling, it is possible to obtain nontrivial results directly from a quantum object.  As an example, let us consider the Jaynes-Cummings model in the ultra-strong coupling regime $g \ge \omega_{0},\epsilon$ where the rotating wave approximation (RWA) is no longer valid
\begin{equation}\label{eq:nonrwa}
H = \omega_{0} a^{\dag} a  + \frac{1}{2}\epsilon\sigma_z+g\left(a^{\dag}+a\right)\left(\sigma_{+}+\sigma_{-}\right).
\end{equation}
Recently, this regime has become of interest \cite{casanova:2010,ashhab:2010,cao:2011} due to the experimental realization of the required large coupling strengths in superconducting circuit devices \cite{forn:2010}.  When the coupling strength $g$ is a significant fraction of the cavity and qubit frequencies, the ground state of the cavity mode, after tracing out the qubit, is no longer the vacuum state.  Instead, the anti-resonant terms proportional to $a^{\dag}\sigma_{+}$ and $a \sigma_{-}$ give rise to an anomalous ground state which, in the large coupling limit $g/\omega_{0} \gg 1$, may be approximated as \cite{nataf:2010,ashhab:2010} 
\begin{equation}\label{eq:approx}
\left|\psi_{g}\right>\simeq \frac{1}{\sqrt{2}}\left[ \left|\alpha\right>_{c}\left|+\right>_{q} - \left|-\alpha\right>_{c}\left|-\right>_{q}\right],
\end{equation}
where the cavity mode is in a Schr\"{o}dinger cat-state with $\left|\alpha\right|\simeq g$.  This ground state can be evaluated by finding the eigenstates and eigenvalues of the Hamiltonian, and can therefore be extracted directly from the \texttt{Qobj} representation of Eq.~(\ref{eq:nonrwa}).  In Fig.~(\ref{fig:nonrwa}) we plot the cavity and qubit occupation numbers for the groundstate of Eq.~(\ref{eq:nonrwa}) as a function of the coupling strength.  Here, the cavity is on resonance with the qubit transition frequency, $\omega_{0}=\epsilon=2\pi$.  In addition, Fig.~\ref{fig:nonrwa} shows the Wigner function for the cavity mode at the largest coupling strength $g=2.5\omega_0$, which is well approximated by Eq.~(\ref{eq:approx}). The 20 lines of QuTiP code used in calculating Fig.~\ref{fig:nonrwa} are given in \ref{sec:nonrwa_code}.

\section{Evolution of open quantum systems}\label{sec:evolution}

The main focus of QuTiP is the time-evolution of open quantum systems. Before we describe how this problem is approached in QuTiP, we give a brief review of the theory of quantum evolution, and the available methods for numerically integrating the equations of motion.

The dynamics of a closed (pure) quantum system is governed by the Schr{\"o}dinger equation
\begin{eqnarray}\label{eq:schrodinger}
i\hbar\frac{\partial}{\partial t}\Psi = \hat H \Psi,
\end{eqnarray}
where $\Psi$ is the wave function, $\hat H$ the Hamiltonian, and $\hbar$ is Planck's constant. In general, the Schr{\"o}dinger equation is a partial differential equation (PDE) where both $\Psi$ and $\hat H$ are functions of space and time. For computational purposes it is useful to expand the PDE in a set of basis functions that span the Hilbert space of the Hamiltonian, and to write the equation in matrix and vector form
\begin{eqnarray}
i\hbar\frac{d}{dt}\left|\psi\right> = H \left|\psi\right>,
\end{eqnarray}
where $\left|\psi\right>$ is the state vector and $H$ is the matrix representation of the Hamiltonian. This matrix equation can, in principle, be solved by diagonalizing the Hamiltonian matrix $H$. In practice, however, it is difficult to perform this diagonalization unless the size of the Hilbert space (dimension of the matrix $H$) is small. Analytically, it is a formidable task to calculate the dynamics for systems with more than two states. If, in addition, we consider dissipation due to the inevitable interaction with a surrounding environment, the computational complexity grows even larger, and we have to resort to numerical calculations in all realistic situations. This illustrates the importance of numerical calculations in describing the dynamics of open quantum systems, and the need for efficient and accessible tools for this task.

While the evolution of the state vector in a closed quantum system is deterministic, open quantum systems are stochastic in nature. The effect of an environment on the system of interest is to induce  stochastic transitions between energy levels, and to introduce uncertainty in the phase difference between states of the system. The state of an open quantum system is therefore described in terms of ensemble averaged states using the density matrix formalism. A density matrix $\rho$ describes a probability distribution of quantum states $\left|\psi_n\right>$, in a matrix representation $\rho = \sum_n p_n \left|\psi_n\right>\left<\psi_n\right|$, where $p_n$ is the classical probability that the system is in the quantum state $\left|\psi_n\right>$. The time evolution of a density matrix $\rho$ is the topic of the remaining portions of this section.

\subsection{Master equation}\label{sec:master}

The standard approach for deriving the equations of motion for a system interacting with its environment is to expand the scope of the system to include the environment. The combined quantum system is then closed, and its evolution is governed by the von Neumann equation
\begin{equation}
\label{eq:neumann_total}
\dot \rho_{\rm tot}(t) = -\frac{i}{\hbar}[H_{\rm tot}, \rho_{\rm tot}(t)],
\end{equation}
the equivalent of the Schr{\"o}dinger equation (\ref{eq:schrodinger}) in the density matrix formalism. Here, the total Hamiltonian 
\begin{equation}
 H_{\rm tot} = H_{\rm sys} + H_{\rm env} + H_{\rm int},
\end{equation}
includes the original system Hamiltonian $H_{\rm sys}$, the Hamiltonian for the environment $H_{\rm env}$, and a term representing the interaction between the system and its environment $H_{\rm int}$. Since we are only interested in the dynamics of the system, we can at this point perform a partial trace over the environmental degrees of freedom in Eq.~(\ref{eq:neumann_total}), and thereby obtain a master equation for the motion of the original system density matrix. The most general trace-preserving and completely positive form of this evolution is the Lindblad master equation for the reduced density matrix $\rho = {\rm Tr}_{\rm env}[\rho_{\rm tot}]$ 
\begin{eqnarray}
\label{eq:master_equation}
\!\!\dot \rho(t) \!\!\!\!&=&\!\!\! -\frac{i}{\hbar}[H(t), \rho(t)]  \nonumber\\
&+&\!\!\!\!\!
\sum_n \frac{1}{2} \left[2 C_n \rho(t) C_n^\dag - \rho(t) C_n^\dag C_n - C_n^\dag C_n \rho(t)\right]\!,\,\,
\end{eqnarray}
where the $C_n = \sqrt{\gamma_n} A_n$ are collapse operators, and $A_n$ are the operators through which the environment couples to the system in $H_{\rm int}$, and $\gamma_n$ are the corresponding rates.  The derivation of Eq.~(\ref{eq:master_equation}) may be found in several sources \cite{lindblad:1976,gardiner:2004,walls:2008}, and will not be reproduced here.  Instead, we emphasize the approximations that are required to arrive at the master equation in the form of Eq.~(\ref{eq:master_equation}), and hence perform a calculation in QuTiP:

\renewcommand{\labelitemi}{}
\begin{itemize}
\item \textbf{Separability:} At $t=0$ there are no correlations between the system and its environment such that the total density matrix can be written as a tensor product $\rho^I_{\rm tot}(0) = \rho^I(0) \otimes \rho^I_{\rm env}(0)$.

\item \textbf{Born approximation:} Requires: (1) that the state of the environment does not significantly change as a result of the interaction with the system;  (2) The system and the environment remain separable throughout the evolution. These assumptions are justified if the interaction is weak, and if the environment is much larger than the system. In summary, $\rho_{\rm tot}(t) \approx \rho(t)\otimes\rho_{\rm env}$.

\item \textbf{Markov approximation:} The time-scale of decay for the environment $\tau_{\rm env}$ is much shorter than the smallest time-scale of the system dynamics $\tau_{\rm sys} \gg \tau_{\rm env}$. This approximation is often deemed a ``short-memory environment'' as it requires that environmental correlation functions decay on a time-scale fast compared to those of the system.

\item \textbf{Secular approximation:} Stipulates that elements in the master equation corresponding to transition frequencies satisfy $|\omega_{ab}-\omega_{cd}| \ll 1/\tau_{\rm sys}$, i.e., all fast rotating terms in the interaction picture can be neglected. It also ignores terms that lead to a small renormalization of the system energy levels. This approximation is not strictly necessary for all master-equation formalisms (e.g., the Block-Redfield master equation), but it is required for arriving at the Lindblad form (\ref{eq:master_equation}) which is used in QuTiP.
\end{itemize}

For systems with environments satisfying the conditions outlined above, the Lindblad master equation (\ref{eq:master_equation}) governs the time-evolution of the system density matrix, giving an ensemble average of the system dynamics. In order to ensure that these approximations are not violated, it is important that the decay rates $\gamma_n$ be smaller than the minimum energy splitting in the system Hamiltonian. Situations that demand special attention therefore include, for example, systems strongly coupled to their environment, and systems with degenerate or nearly degenerate energy levels. 

\begin{figure}[t]
\begin{center}
\includegraphics[width=9.5cm]{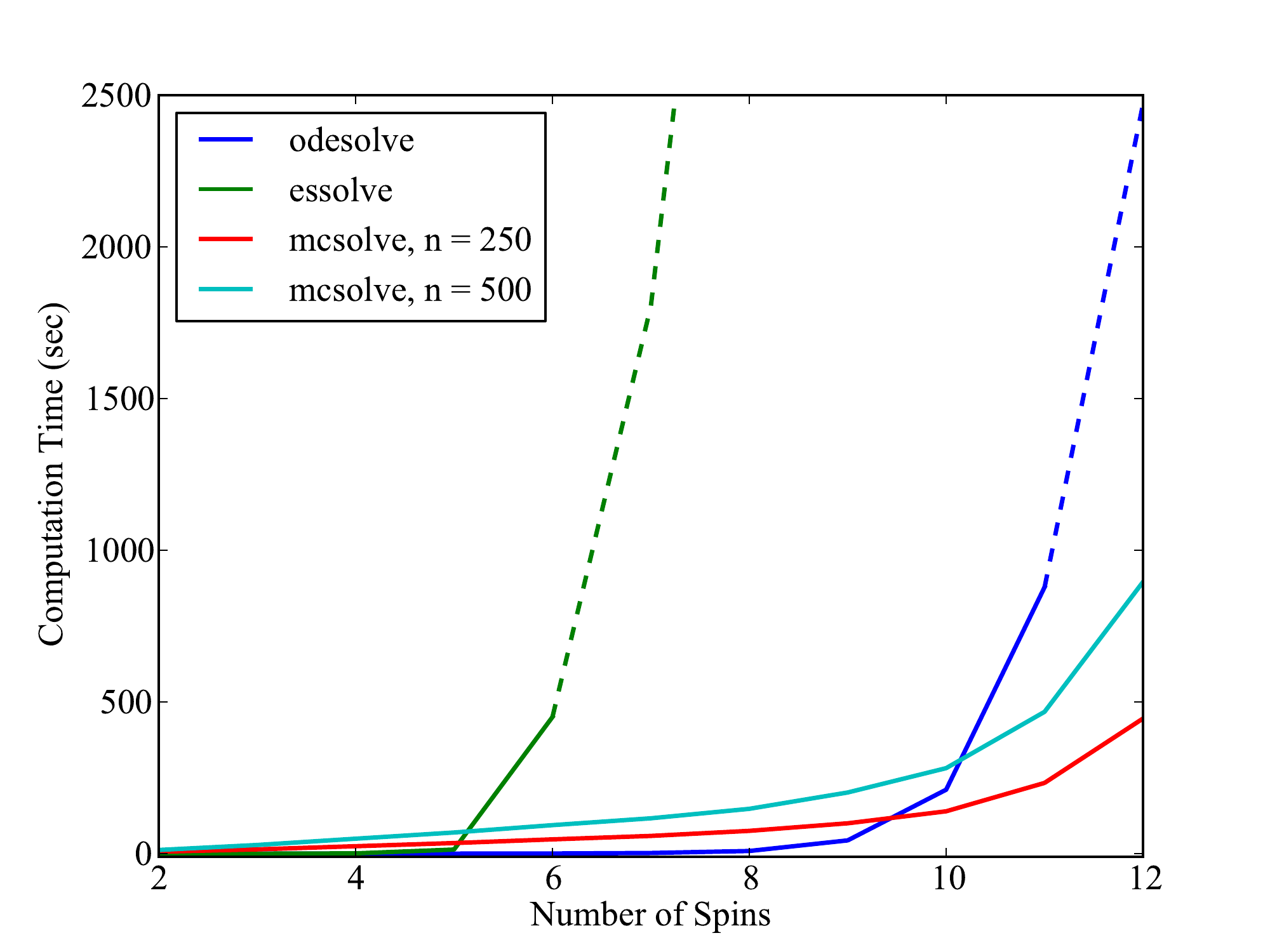}
\caption{(Color) The time required to evolve the Heisenberg spin-chain, Eq.~(\ref{eq:spinchain}), as a function of the system size $2^{M}$ where $M$ is the number of spins, using the master equation ODE solver \texttt{odesolve} (blue), diagonalization via \texttt{essolve} (green), and the Monte-Carlo solver \texttt{mcsolve} with 250 (red) and 500 (cyan) trajectories, respectively.  The dashed lines give the estimated calculation times extrapolated from the data when the simulation could no longer fit in the computers memory (odesolve), or the calculation became intractable (essolve).  Here, the spin parameters are assumed to be identical with $h=2\pi$ and $J_{x}=J_{y}=J_{z}=0.1\times2\pi$.  Likewise, each spin has a dephasing rate given by $\gamma=0.01$.  The initial state is given by $|\psi(0)\rangle=|1\rangle_{1}|0\rangle_{2}\dots|0\rangle_{M}$.  Calculations were performed on a 2.8~Ghz quad-core computer with 24~Gb of memory.}
\label{fig:solver-performance}
\end{center}
\end{figure}

In QuTiP there are two solvers that calculate the time evolution according to Eq.~(\ref{eq:master_equation}): \texttt{odesolve} numerically integrates the set of coupled ordinary differential equations (ODEs), and \texttt{essolve} which employs full diagonalization. The \texttt{odesolve} and \texttt{essolve} solvers both take the same set of input parameters (as exemplified in Sec.~\ref{sec:numerical}) and can easily be substituted for each other in a QuTiP program. For a quantum system with $N$ states, the number of elements in the density matrix is $N^2$, and solving the master equation by numerical integration or diagonalization involves of use of superoperators of size $N^2 \times N^2$. In the sparse matrix format, not all of the $N^4$ elements need to be stored in the memory.  However, the time required to evolve a quantum system according to the master equation still increases rapidly as a function of the system size. Consequently, the master equation solvers are practical only for relatively small systems: $N \lesssim1000$, depending on the details of the problem. In Fig.~\ref{fig:solver-performance} we show the scaling of the elapsed time for a typical simulation, here chosen to be the Heisenberg spin-chain 
\begin{eqnarray}\label{eq:spinchain}
H = - \frac{1}{2} \sum_{n}^{M} h^{n} \sigma_{z}^{n}-\frac{1}{2} \sum_{n}^{M-1} \left[ J_{x}^{n} \sigma_{x}^{n} \sigma_{x}^{n+1}\right. \\ \nonumber
\left.+ J_{y}^{n} \sigma_{y}^{n} \sigma_{y}^{n+1}+ J_{z}^{n} \sigma_{z}^{n} \sigma_{z}^{n+1}\right],
\end{eqnarray}
as a function of the size of the Hilbert space, for the two master equation solvers, as well as for two realizations of the Monte-Carlo solver \texttt{mcsolve} described the following section.  In general, the exact time required to evolve a system depends on the details of the problem, but the scaling with system size is rather generic.  The Monte-Carlo solver has superior scaling properties compared to the master-equation solvers, but due to the overhead from stochastic averaging, it is only for systems with a Hilbert space dimension around $\sim1000$ that the Monte-Carlo solvers outperform the master equation.

\subsection{Monte-Carlo trajectories}\label{sec:monte}
Where as the density matrix formalism describes the ensemble average over many identical realizations of a quantum system, the Monte-Carlo (MC), or quantum-jump approach \cite{plenio:1998} to wave function evolution, allows for simulating an individual realization of the system dynamics.  Here, the environment is continuously monitored, resulting in a series of quantum jumps in the system wave function, conditioned on the increase in information gained about the state of the system via the environmental measurements \cite{haroche:2006}.  In general, this evolution is governed by the Schr\"{o}dinger equation (\ref{eq:schrodinger}) with a \textit{non-Hermitian} effective Hamiltonian  
\begin{equation}\label{eq:heff}
H_{\rm eff}=H_{\rm sys}-\frac{i\hbar}{2}\sum_{i}C^{\dag}_{n}C_{n},
\end{equation}
where again, the $C_{n}$ are collapse operators, each corresponding to a separate irreversible process with rate $\gamma_{n}$.  Here, the strictly negative non-Hermitian portion of Eq.~(\ref{eq:heff}) gives rise to a reduction in the norm of the wave function, that to first-order in a small time $\delta t$, is given by $\left<\psi(t+\delta t)|\psi(t+\delta t)\right>=1-\delta p$ where
\begin{equation}\label{eq:jump}
\delta p =\delta t \sum_{n}\left<\psi(t)|C^{\dag}_{n}C_{n}|\psi(t)\right>,
\end{equation}
and $\delta t$ is such that $\delta p \ll 1$.  With a probability of remaining in the state $\left|\psi(t+\delta t)\right>$ given by $1-\delta p$, the corresponding quantum jump probability is thus Eq.~(\ref{eq:jump}).  If the environmental measurements register a quantum jump, say via the emission of a photon into the environment \cite{guerlin:2007}, or a change in the spin of a quantum dot \cite{vamivakas:2010}, the wave function undergoes a jump into a state defined by projecting $\left|\psi(t)\right>$ using the collapse operator $C_{n}$ corresponding to the measurement
\begin{equation}\label{eq:project}
\left|\psi(t+\delta t)\right>=C_{n}\left|\psi(t)\right>/\left<\psi(t)|C_{n}^{\dag}C_{n}|\psi(t)\right>^{1/2}.
\end{equation}
If more than a single collapse operator is present in Eq~(\ref{eq:heff}), the probability of collapse due to the $i\mathrm{th}$-operator $C_{i}$ is given by 
\begin{equation}\label{eq:pcn}
P_{i}(t)=\left<\psi(t)|C_{i}^{\dag}C_{i}|\psi(t)\right>/\delta p.
\end{equation}

Evaluating the MC evolution to first-order in time is quite tedious.  Instead, QuTiP uses the following algorithm to simulate a single realization of a quantum system \cite{dalibard:1992,dum:1992,molmer:1993}.  Starting from a pure state $\left|\psi(0)\right>$:

\begin{itemize}
\item \textbf{I:} Choose a random number $r$ between zero and one, representing the probability that a quantum jump occurs.  

\item \textbf{II:} Integrate the Schr\"{o}dinger equation (\ref{eq:schrodinger}), using the effective Hamiltonian (\ref{eq:heff}) until a time $\tau$ such that the norm of the wave function satisfies $\left<\psi(\tau)\right. \left|\psi(\tau)\right>=r$, at which point a jump occurs.

\item \textbf{III:}  The resultant jump projects the system at time $\tau$ into one of the renormalized states given by Eq.~(\ref{eq:project}).  The corresponding collapse operator $C_{n}$ is chosen such that $n$ is the smallest integer satisfying
\begin{equation}\label{eq:mc3}
\sum_{i=1}^{n} P_{n}(\tau) \ge r
\end{equation}
where the individual $P_{n}$ are given by Eq.~(\ref{eq:pcn}).  Note that the left hand side of Eq.~(\ref{eq:mc3}) is, by definition, normalized to unity.

\item \textbf{IV:}  Using the renormalized state from step III as the new initial condition at time $\tau$, draw a new random number, and repeat the above procedure until the final simulation time is reached.
\end{itemize}
\renewcommand{\labelitemi}{$\bullet$}

\subsubsection{Example: Single-photon cavity decay}

\begin{figure}[t]
\begin{center}
\includegraphics[width=7.25cm]{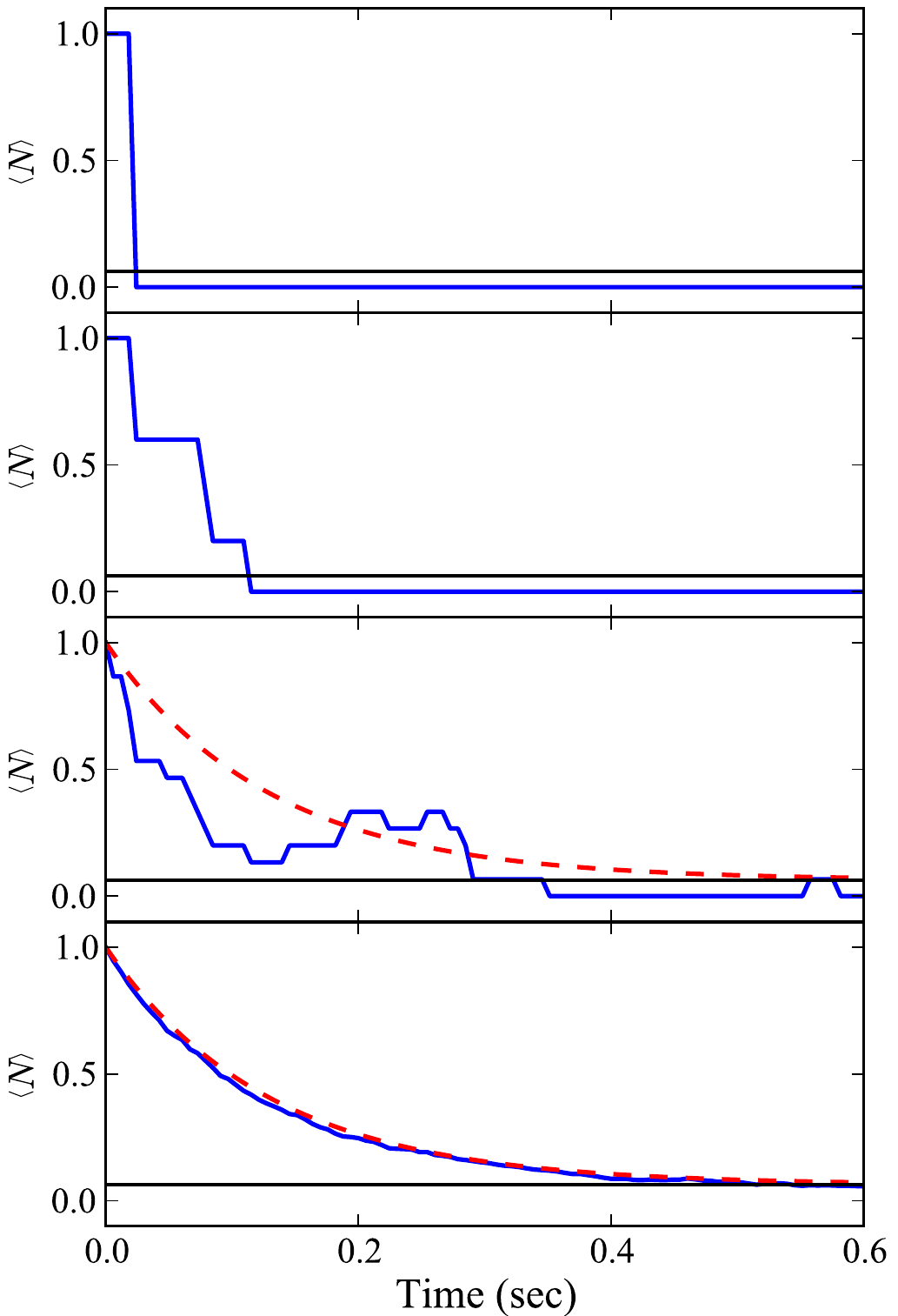}
\caption{(Color) (a) Monte-Carlo simulation showing the number operator expectation value for a single trajectory (blue) in the decay of a single-photon Fock state from a cavity coupled to a thermal environment, as demonstrated experimentally in Ref.~\cite{gleyzes:2007}.  Here, the average photon number for the thermal environment is $n=0.063$ (black), while the decay rate $\kappa=1/T_{c}$ is given by the cavity ring-down time $T_{c}=0.129$. (b-d) Averages of 5, 15, and 904 trajectories showing ensemble averaging toward the master equation solution (dashed-red).}
\label{fig:decay}
\end{center}
\end{figure}

As an illustrative example, let us consider the evolution of a single-photon cavity Fock state in a non-zero thermal environment \cite{gleyzes:2007}.  The evolution of the wave function $\left|\psi(0)\right>=\left|1\right>$ is governed by the effective Hamiltonian ($\hbar=1$)
\begin{equation}\label{eq:heffth}
H_{\rm eff}=\omega_{c}a^{\dag}a-\frac{1+\left<n \right>_{\rm th}}{2}i\kappa a^{\dag}a-\frac{\left<n \right>_{\rm th}}{2}i\kappa a a^{\dag},
\end{equation}
with cavity frequency $\omega_{c}$, cavity decay rate $\kappa$, and where $\left<n\right>_{\rm th}$ is the steady state thermal occupation number.  While the first term in Eq.~(\ref{eq:heff}) is responsible for the standard unitary evolution of the cavity mode, the second and third terms give rise to random quantum jumps to lower and higher cavity photon numbers, respectively.  When a jump occurs, the wave function of the system is projected into a state corresponding to the collapse operator $C_{1}=\sqrt{(1+\left< n \right>_{\rm th})\kappa}a$, yielding a decrease in the cavity occupation number, or $C_{2}=\sqrt{\left< n \right>_{\rm th}\kappa}a^{\dag}$, which results in an increase.  Here, the relative ratio of jumps corresponding to an increase in the cavity occupation number to those for decay is determined by the magnitude of $\left< n \right>_{\rm th}$. A single realization of this evolution, showing a lone quantum jump, is presented in Fig.~\ref{fig:decay}a.   

\begin{figure}[t]
\begin{center}
\includegraphics[width=8.5cm]{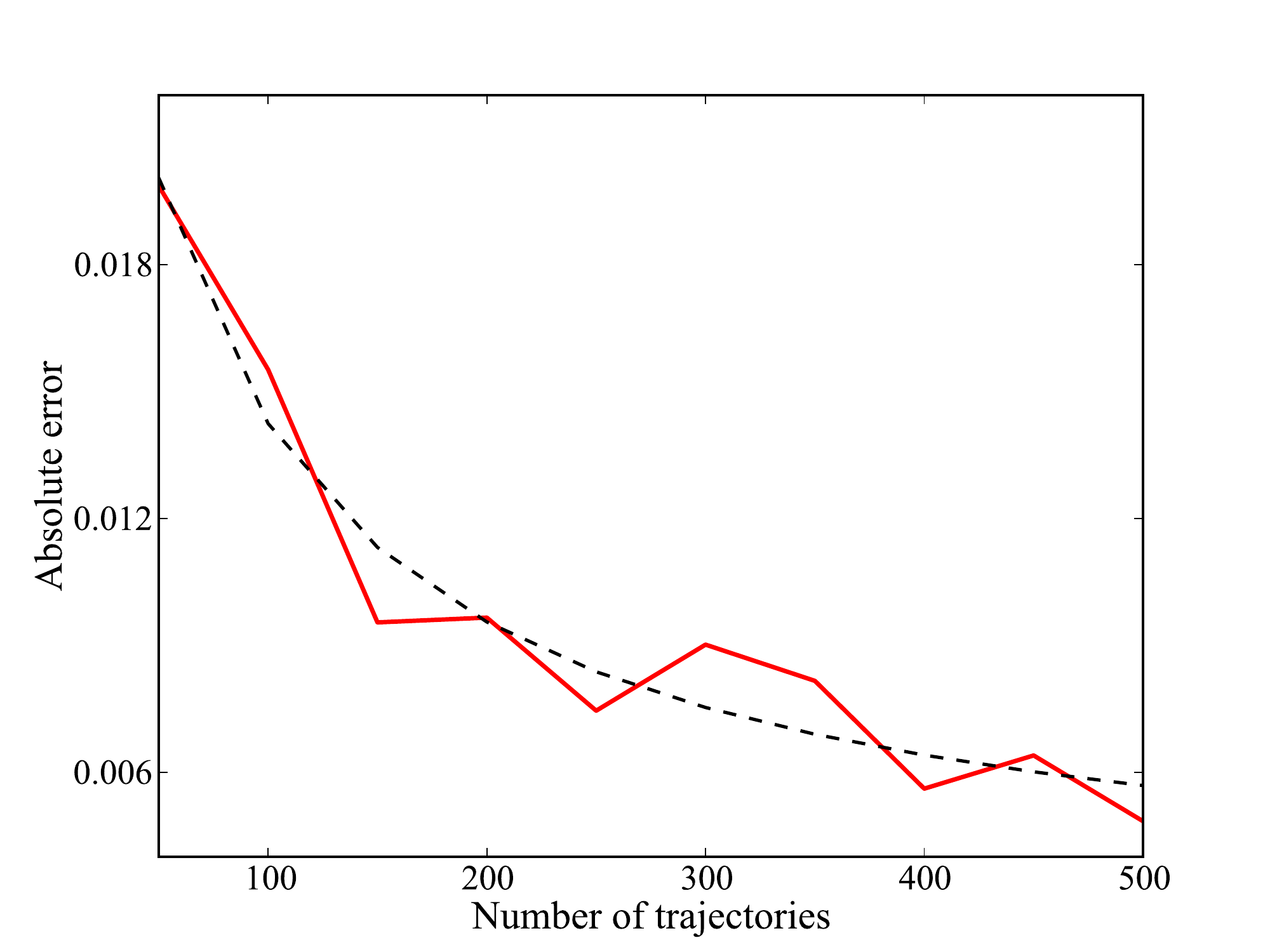}
\caption{(Color) Scaling of the sum of the absolute deviation per time step, averaged over $10$ simulations, for expectation values calculated with the Monte-Carlo solver and compared to the ME solution, as a function of the number of trajectories. The error measure is calculated by integrating the absolute errors over the entire evolution time, and normalized by the number of time steps. The acceptable level of error varies from case to case, but in general the error measure should be much smaller than unity.  A $1/m$ fit to the data (dashed) showing the predicted convergence rate is also presented.}
\label{fig:sse}
\end{center}
\end{figure}

In addition to single quantum systems, when averaged over a sufficiently large number of identical system realizations, the MC method leads to the same evolution equation as the Lindblad master equation (\ref{eq:master_equation}) for a pure state density matrix \cite{plenio:1998,haroche:2006}.  Therefore, the MC method may be used in any situation where the Lindblad master equation is valid, as discussed in Sec.~\ref{sec:master}.  However, for large quantum systems with Hilbert space dimension $N\gg 1$, the MC method is vastly more efficient than simulating the full density matrix given that only $N$ elements are required to simulate a wave function, as opposed to the $N^{2}$ elements necessary in the ME approach.  Although multiple trajectories are required, convergence to the ME result scales as $m^{-1}$ \cite{walls:2008}, where $m$ is the number of trajectories simulated.  In typical situations, between 250 and 500 trajectories are sufficient for errors smaller than a few percent, see Fig.~\ref{fig:sse}.  In Fig.~\ref{fig:decay} we show the convergence of the MC simulation to the ME solution for the single-photon cavity example as the number of trajectories averaged over is increased.

\section{Numerical calculations}\label{sec:numerical}

In this section we illustrate, via a number of examples, how quantum dynamical calculations are carried out using the QuTiP framework.
The typical workflow for performing a simulation with qutip is: 
\renewcommand{\labelitemi}{}
\begin{itemize}
\item \textbf{I:}  Define the parameters that characterize the system and environment (if applicable).
\item \textbf{II:}  Create \texttt{Qobj} class instances representing the Hamiltonian and initial state of the system.
\item \textbf{III:}  For dissipative systems, define the collapse operators as \texttt{Qobj} objects.
\item \textbf{IV:}  Evolve the system with a choice of evolution algorithm and output (e.g., operator expectation values).
\item \textbf{V:}  Post-process and visualize the data. 
\end{itemize}
\renewcommand{\labelitemi}{$\bullet$}
Using the quantum object described in Sec.~\ref{sec:framework}, and the solvers for the time-evolution of quantum systems described in Sec.~\ref{sec:evolution}, we can explore a diverse set of problems. The examples presented here are selected because they illustrate different features of QuTiP with a minimum of complexity.

\subsection{Fidelity of a two-qubit gate subject to noise}\label{sec:numerical_coupled_qubits}

To introduce how the evolution of a dynamical quantum system is calculated using QuTiP, let us consider a simple system comprised of two qubits that, during a time $T=\pi/4g$, are subject to the coupling Hamiltonian
\begin{equation}
\label{eq:hamiltonian_iswap}
H = g \left(\sigma_x\otimes\sigma_x + \sigma_y\otimes\sigma_y\right),
\end{equation}
where $g$ is the coupling strength. Under ideal conditions this coupling realizes the $i$-SWAP gate between the two qubit states \cite{schuch:2003, steffen:2006}. This can readily be seen by evolving any initial state for the time $T$, and comparing the final state with the corresponding $i$-SWAP transformed initial state. We shall assume that the qubits are coupled with their surrounding environments, resulting in qubit energy relaxation and dephasing. 

\begin{figure}[t]
\begin{center}
\includegraphics[width=8.5cm]{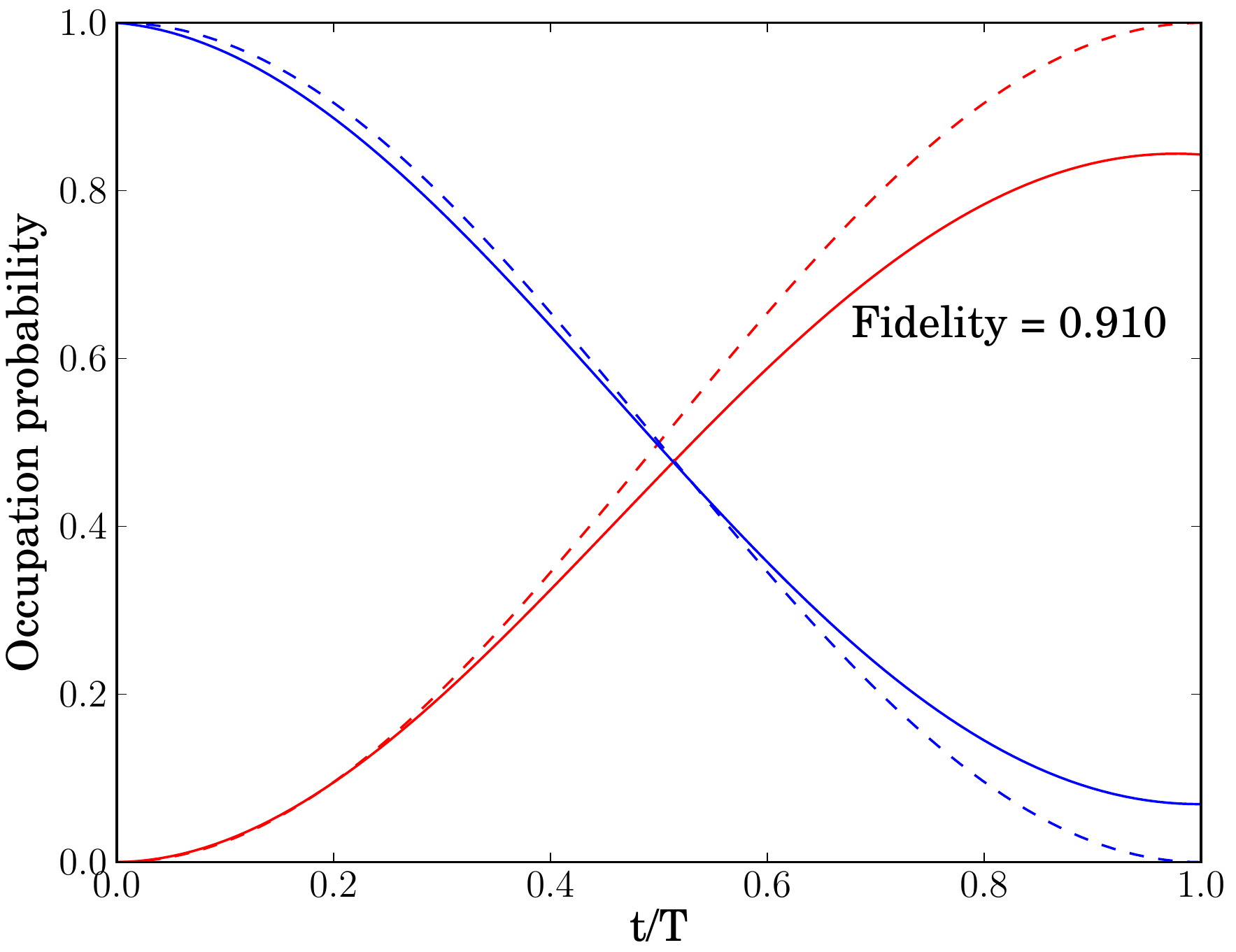}
\caption{(Color) The time-evolution of a two-qubit system described by the Hamiltonian Eq.~(\ref{eq:hamiltonian_iswap}), which at time $t/T = 1$ and ideal conditions (dashed lines) transforms the qubit states in accordance with the $i$-SWAP gate. With qubit relaxation and dephasing (solid lines), the end-result deviates from the ideal $i$-SWAP gate. For these particular parameters ($g=2\pi$, $\Gamma_1=0.75$, $\Gamma_2=0.5$), the fidelity of the dissipative gate is 91\%.}
\label{fig:iswap}
\end{center}
\end{figure}

Following the workflow outlined in the previous section, the QuTiP code for this problem can be organized in the following manner. First, define the numerical constants in the problem. For brevity, the code for this step has been omitted. Next, the \texttt{Qobj} instances for the Hamiltonian and the initial state may be defined as
\begin{footnotesize}
\begin{verbatim}
H = g * (tensor(sigmax(), sigmax()) +
         tensor(sigmay(), sigmay()))

psi0 = tensor(basis(2,1), basis(2,0))
\end{verbatim}
\end{footnotesize}
To model qubit relaxation and dephasing, we define a list of collapse operators that later will be passed on to the ODE solver. For each qubit we append its associated collapse operator to this list (here called \texttt{c\_op\_list})
\begin{footnotesize}
\begin{verbatim}        
sm1 = tensor(sigmam(), qeye(2))
sz1 = tensor(sigmaz(), qeye(2))
c_op_list.append(sqrt(g1 * (1+nth)) * sm1)
c_op_list.append(sqrt(g1 * nth) * sm1.dag())
c_op_list.append(sqrt(g2) * sz1)
\end{verbatim}
\end{footnotesize}
where the parameter \texttt{nth} is the number of environmentally-induced thermal excitations in the steady state. The collapse operators containing $\sigma_-$ and $\sigma_-^\dag$ describe the qubit relaxation and excitation with the rates \texttt{g1 * (1+nth)} and \texttt{g1 * nth}, respectively, and the collapse operator $\sigma_z$ models qubit dephasing. These lines of codes are repeated for the second qubit, with the appropriate change in the definition of the operators (i.e., the arguments in the \texttt{tensor} function are switched).

At this point we are ready to let QuTiP calculate the time-evolution of the system. In the following example we use the master equation ODE solver \texttt{odesolve}. In addition to the Hamiltonian, initial state, and the list of collapse operator, we pass a list \texttt{tlist} to the solver that contains the times at which we wish to evaluate the density matrix.
\begin{footnotesize}
\begin{verbatim}
tlist     = linspace(0, T, 100)
rho_list  = odesolve(H, psi0, tlist, c_op_list, [])
rho_final = rho_list[-1]
\end{verbatim}
\end{footnotesize}
If the last parameter is empty, as in this example, all QuTiP time-evolution solvers return the full density matrix (or state vector) corresponding to the times in  \texttt{tlist}. Alternatively, a list of operators may be passed as last argument to the solver, in which case it will return the corresponding expectation values.

Given the output of density matrices, we may now calculate the corresponding expectation values for selected quantum operators. For example, to calculate the excitation probability of the two qubits as a function of time, we may use the QuTiP function \texttt{expect}
\begin{footnotesize}
\begin{verbatim}
n1 = expect(sm1.dag() * sm1, rho_list)
n2 = expect(sm2.dag() * sm2, rho_list)
\end{verbatim}
\end{footnotesize}
Here, \texttt{n1} and \texttt{n2} are now real \textit{NumPy} arrays of expectation values, suitable for plotting or saving to file.

Finally, to quantify the difference between the lossy $i$-SWAP gate and its ideal counterpart, we calculate the fidelity \texttt{fidelity}
\begin{footnotesize}
\begin{verbatim}        
U = (-1j * H * pi / (4*g)).expm()
psi_ideal = U * psi0
rho_ideal = psi_ideal * psi_ideal.dag()
f = fidelity(rho_ideal, rho_final). 
\end{verbatim}
\end{footnotesize}

The results are shown in Fig.~\ref{fig:iswap}, where the expectation values for the two qubits, as a function of time, is plotted both with and without dissipation. The full code for this example is listed in \ref{sec:iswap_code}.

\subsection{Jaynes-Cumming model}\label{sec:numerical_jc}

\begin{figure}[t]
\begin{center}
\includegraphics[width=8.5cm]{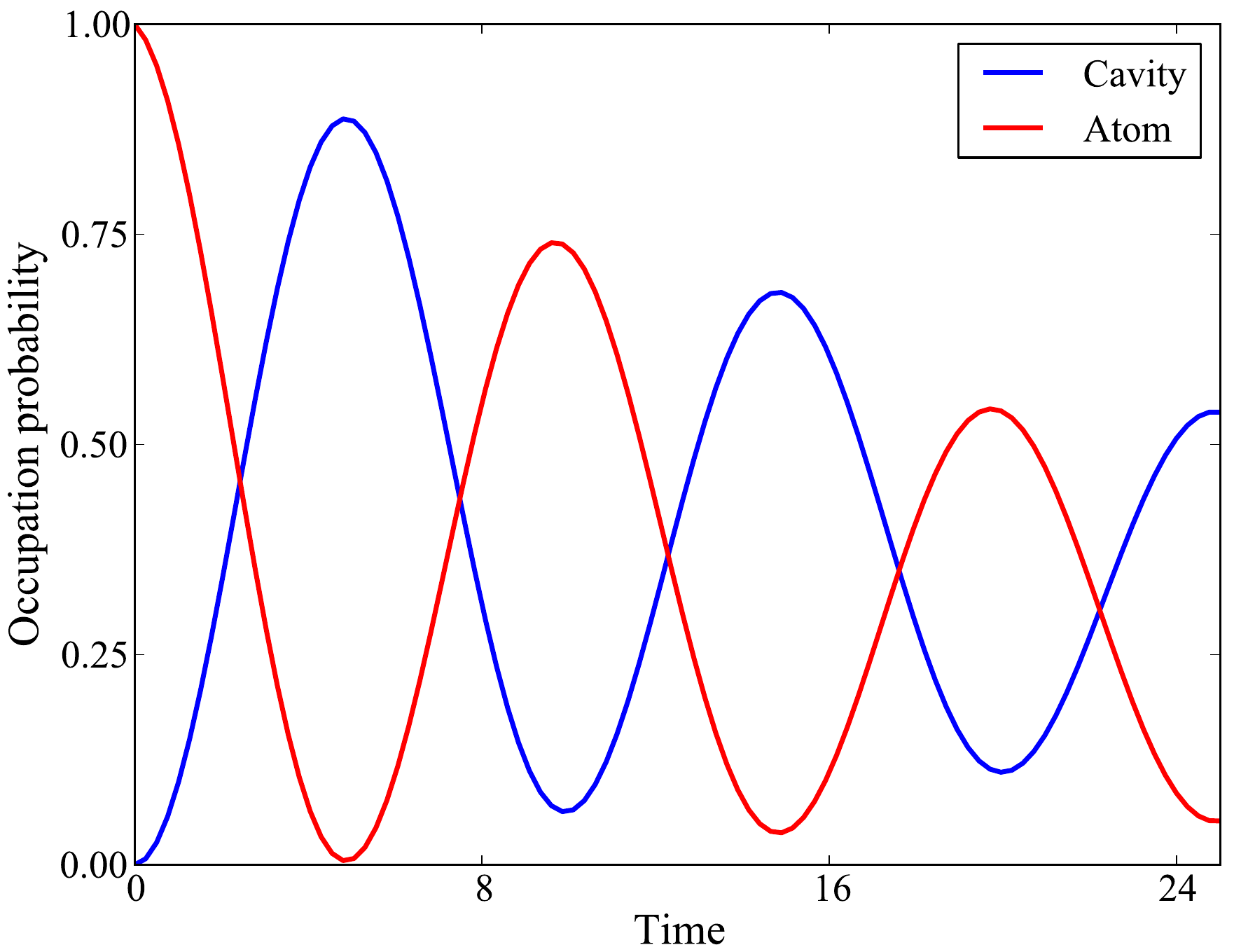}
\caption{(Color) Evolution of the Jaynes-Cummings Hamiltonian (\ref{eq:hamiltonian_jc}) in a thermal environment characterized by $\left< n \right>_{\rm th}=0.75$.  Initially, only the atom is excited, but the atom-cavity coupling results in a coherent energy transfer between the two systems, a phenomenon known as vacuum Rabi oscillations. Here, the atom and the cavity are resonant, $\omega_0 = \epsilon = 2\pi$, the coupling strength $g/\omega_0=0.05$, and the atom and cavity relaxation rates are $\gamma/(\omega_0/2\pi)=0.05$ and $\kappa/(\omega_0/2\pi)=0.005$, respectively.}
\label{fig:jc}
\end{center}
\end{figure}

The same method used in the previous section can calculate the dynamics of the Jaynes-Cumming model Eq.~(\ref{eq:hamiltonian_jc}). Only the definitions of the Hamiltonian, initial state, and collapse operators need to be changed to solve this problem. For the Jaynes-Cumming model, the Hamiltonian and a possible initial state were given in Sec.~\ref{sec:framework}, and we need only to define collapse operators before the system can be evolved using one of the QuTiP solvers. The cavity and the atom relaxation rates are $\kappa$ and $\Gamma$, respectively.  In this example, only the cavity is coupled to an environment with Boltzmann occupation number $n_{\rm th}$.  We can write the collapse operators for the cavity 
\begin{footnotesize}
\begin{verbatim}
a = tensor(destroy(N), qeye(2))
c_ops.append(sqrt(kappa * (1+n_th)) * a)
c_ops.append(sqrt(kappa * n_th) * a.dag())
\end{verbatim}
\end{footnotesize}
and for the atom
\begin{footnotesize}
\begin{verbatim}
sm = tensor(qeye(N), destroy(2))
c_ops.append(sqrt(gamma)* sm)
\end{verbatim}
\end{footnotesize}
Instead of having the solver return the state, as in Sec.~\ref{sec:numerical_coupled_qubits}, we can request that the expectation values for a list of operators be directly calculated at each time step. In this Jaynes-Cumming problem we are interested in the excitation number of the cavity and the atom, and as such can then define a list of expectation value operators
\begin{footnotesize}
\begin{verbatim}
expt_ops = [a.dag() * a, sm.dag() * sm]
\end{verbatim}
\end{footnotesize}
which may be passed as last argument to, for example, the \texttt{odesolve} solver.
\begin{footnotesize}
\begin{verbatim}
tlist     = linspace(0, 10, 100)
expt_list = odesolve(H, psi0, tlist, c_ops, expt_ops)
\end{verbatim}
\end{footnotesize}
The solver then returns a {\it NumPy} array \texttt{expt\_list} of expectation values.  The result of this calculation is shown in Fig.~\ref{fig:jc}, and the complete code is listed in \ref{sec:jc_code}.

\subsection{Trilinear Hamiltonian}\label{sec:tri}
To demonstrate the QuTiP Monte-Carlo (MC) solver, we consider the trilinear Hamiltonian that, in the interaction frame, may be written as
\begin{equation}\label{eq:trilinear}
H=i\hbar K\left(ab^{\dag}c^{\dag}-a^{\dag}bc\right)
\end{equation}
consisting of three harmonic oscillator modes conventionally labeled pump $(a)$, signal $(b)$ and idler $(c)$ respectively, with the frequency relation, $\omega_{a}=\omega_{b}+\omega_{c}$ and coupling constant $K$. This Hamiltonian is the full quantum generalization of the parametric amplifier \cite{mollow:1967} describing several quantum optics processes, including frequency conversion \cite{walls:1970}, the interaction of two-level atoms with a single mode resonant EM field \cite{tavis:1968}, and the modeling of Hawking radiation from a quantized black hole \cite{nation:2010}.  Here we suppose the pump mode is initially in a coherent state, while the signal and idler modes are in the ground state 
\begin{equation}\label{eq:triinit}
\left|\psi(0)\right>=\left|\alpha\right>_{a}\left|0\right>_{b}\left|0\right>_{c}.  
\end{equation}
As a system comprised of three harmonic modes, this model readily lends itself to MC simulation since the Hilbert space dimensionality increases exponentially with the number of initial excitations in the system $\left<N(0)\right>_{a}=\left|\alpha\right|^{2}$.  For example, to accurately model an initial pump mode coherent state with $\left|\alpha\right|^{2}=10$ requires $\sim 17$ states, suggesting a minimum Hilbert space dimensionality of $17^{3}=4913$ for simulating Eq.~(\ref{eq:trilinear}); a value five times larger than what can typically be efficiently calculated using the \texttt{odesolve} or \texttt{eseries} solvers (see Fig.~\ref{fig:solver-performance}). 

\begin{figure}[t]
\begin{center}
\includegraphics[width=8.2cm]{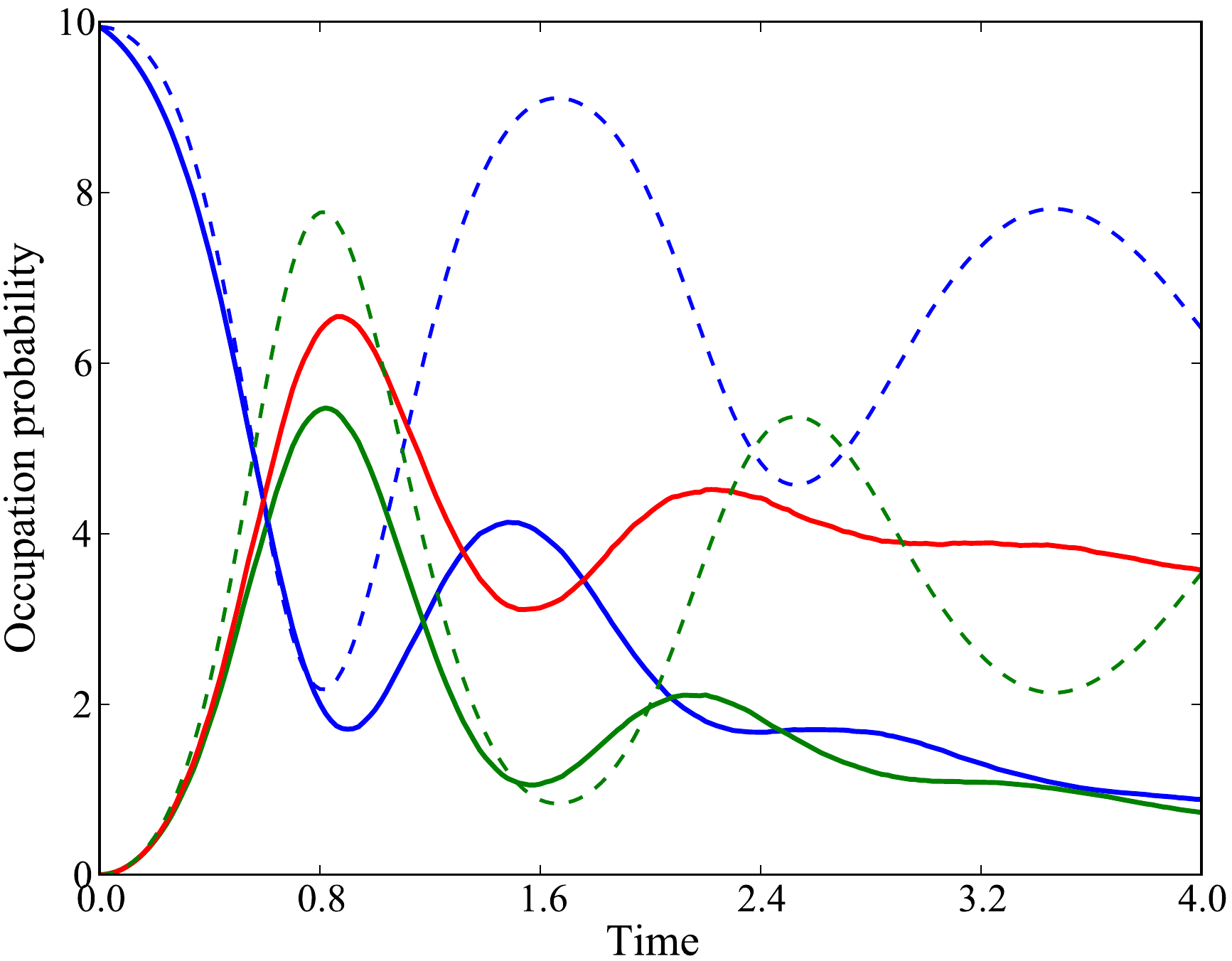}
\caption{(Color) Occupation numbers for the three modes of the trilinear Hamiltonian (\ref{eq:trilinear}), averaged over 1000 trajectories, for an initial state Eq.~(\ref{eq:triinit}) with $\alpha=\sqrt{10}$.  In this simulation, the environment is assumed to be at zero temperature. The pump (blue), signal (green), and idler (red) mode damping rates are $\gamma_{a}=0.2$, $\gamma_{b}=0.8$, and $\gamma_{c}=0.2$, respectively.  The closed system evolution (dashed-colors) is also presented, where the idler mode is omitted as its evolution is identical to that of the signal.}
\label{fig:trilinear}
\end{center}
\end{figure}

In QuTiP, the Hamiltonian (\ref{eq:trilinear}) may be expressed as ($K$=1)
\begin{footnotesize}
\begin{verbatim}
H=1j*(a*b.dag()*c.dag()-a.dag()*b*c),
\end{verbatim}
\end{footnotesize}
with the destruction operators for the pump, signal, and idler modes, $a$, $b$ and $c$ respectively, created via the tensor product
\begin{footnotesize}
\begin{verbatim}
a=tensor(destroy(N),qeye(N),qeye(N))
b=tensor(qeye(N),destroy(N),qeye(N))
c=tensor(qeye(N),qeye(N),destroy(N)).
\end{verbatim}
\end{footnotesize}
In addition, we may define number operator expectation values and collapse operators in the same manner as previous examples
\begin{footnotesize}
\begin{verbatim}
num0,num1,num2=[a0.dag()*a0,a1.dag()*a1,a2.dag()*a2]
C0,C1,C2=[sqrt(2.0*g0)*a0,sqrt(2.0*g1)*a1,sqrt(2.0*g2)*a2]
\end{verbatim}
\end{footnotesize}
\texttt{mcsolve} takes the same input arguments as \texttt{odesolve}, save for an additional argument necessary to specify the number of MC trajectories to simulate
\begin{footnotesize}
\begin{verbatim}
ntraj=500
\end{verbatim}
\end{footnotesize}
In Fig.~\ref{fig:trilinear} we plot the expectation values for the three modes of the trilinear Hamiltonian (\ref{eq:trilinear}), with corresponding damping rates, $\gamma_{a}=0.2$, $\gamma_{b}=0.8$, $\gamma_{c}=0.2$, for the initial state given by Eq.~(\ref{eq:triinit}) with $\alpha=\sqrt{10}$
\begin{footnotesize}
\begin{verbatim}
psi0=tensor(coherent(N,sqrt(10)),basis(N,0),basis(N,0))
avgs=mcsolve(H,psi0,tlist,ntraj,[C0,C1,C2],[num0,num1,num2])
\end{verbatim}
\end{footnotesize}

Had we not defined any collapse operators, the evolution calculated by \texttt{mcsolve} reduces to the Schr\"{o}dinger equation (\ref{eq:schrodinger}).  This evolution is also presented in Fig.~(\ref{fig:trilinear}). The underlying QuTiP code may be found in \ref{sec:trilinear_code}.

\subsection{Landau-Zener transitions}\label{sec:landau}

Landau-Zener transitions \cite{shevchenko:2010} are an interesting problem that involves a quantum two-level system with a time-dependent energy splitting. The Hamiltonian is
\begin{equation}
\label{eq:lz-hamiltonian}
H(t) = \frac{\Delta}{2} \sigma_x + \frac{vt}{2} \sigma_z,
\end{equation}
where $\Delta$ is the tunneling rate, and $v$ is the rate of change in the bare qubit energy splitting. 
The Landau-Zener transition theory analytically describes how the final state at $t\rightarrow\infty$ is related to the initial state at $t\rightarrow-\infty$. In particular, the probability of an adiabatic transition from $\left|1\right>$ to $\left|0\right>$ is given by the Landau-Zener formula
\begin{equation}
\label{eq:lz}
P = 1 - \exp\left(-\frac{\pi\Delta^2}{2v}\right).
\end{equation}
Using QuTiP, we can easily integrate the system dynamics numerically, and obtain the state of the system for any intermediate value of $t$.

\begin{figure}[t]
\begin{center}
\includegraphics[width=8.5cm]{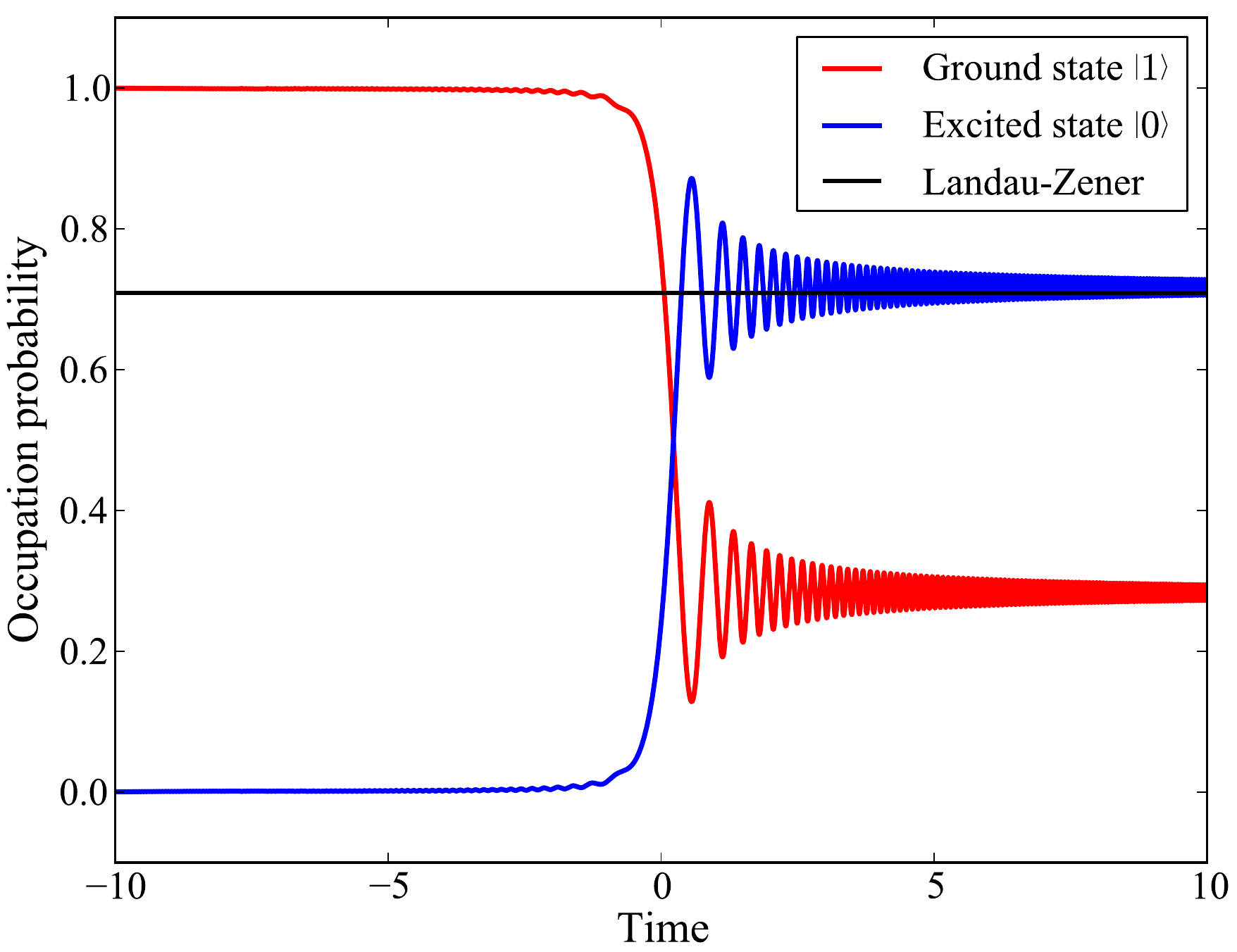}
\caption{(Color) The occupation probability of the $\left|1\right>$ (red curve) and $\left|0\right>$ (blue curve) states of a quantum two-level system throughout a Landau-Zener transition. The solid black line is the final state according to the Landau-Zener formula Eq.~(\ref{eq:lz}). The parameters used in this calculation are $\Delta = 0.5 \times 2\pi$ and $v = 2.0 \times 2\pi$.}
\label{fig:lz_occupation}
\end{center}
\end{figure}

\begin{figure}[t]
\begin{center}
\includegraphics[width=6.5cm]{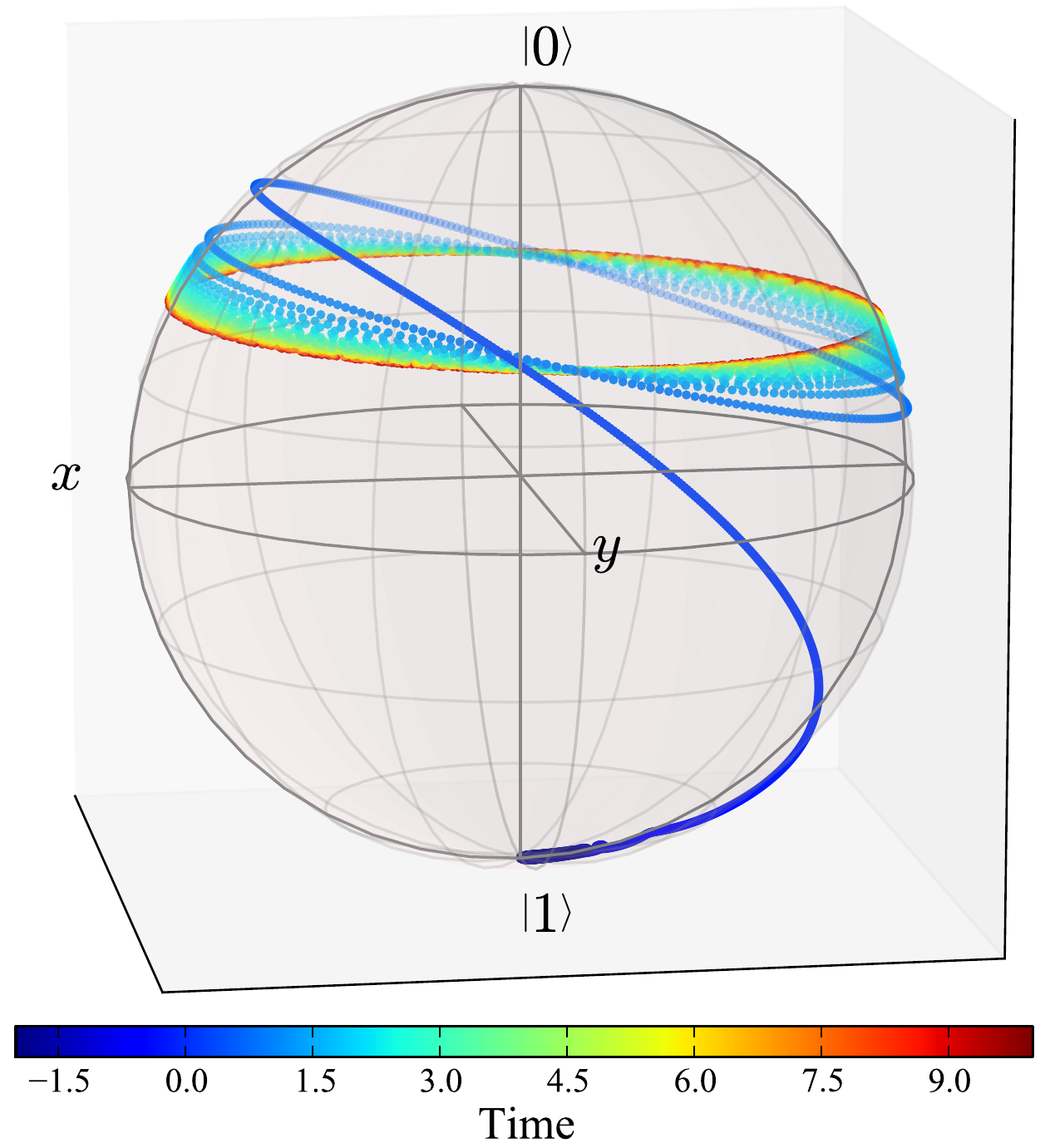}
\caption{(Color) Bloch sphere representation of the Landau-Zener transition presented in Fig.~\ref{fig:lz_occupation}.  Here, the color of the data points corresponds to the evolution time.}
\label{fig:lz_bloch}
\end{center}
\end{figure}

The Landau-Zener problem differs from the previous examples in that the Hamiltonian is explicitly time-dependent. The QuTiP solvers \texttt{odesolve} and \texttt{mcsolve} both support time-dependent Hamiltonians. In order to specify an arbitrary time-dependent Hamiltonian, a callback function may be passed as the first argument to the time-evolution solvers (in place of the \texttt{Qobj} instance that normally represents the Hamiltonian). The callback function is expected to return the value of the Hamiltonian at the given point in time \texttt{t}, which is the first argument to the callback function.
\begin{footnotesize}
\begin{verbatim}        
def hamiltonian_t(t, args):
    H0 = args[0]
    H1 = args[1]
    return H0 + t * H1
\end{verbatim}
\end{footnotesize}
In addition, a second argument passed to the callback function is a user-defined list of parameters. For performance reasons, it is appropriate to let this list contain pre-calculated \texttt{Qobj} instances for the constant parts of the Hamiltonian. For the Landau-Zener problem this corresponds to
\begin{footnotesize}
\begin{verbatim}        
H0 = delta/2.0 * sigmax()
H1 =     v/2.0 * sigmaz()
H_args = (H0, H1)
\end{verbatim}
\end{footnotesize}
The list of arguments for the Hamiltonian callback function is then passed on to the time-evolution solver (as the very last argument), along with the callback function itself (as first argument).
\begin{footnotesize}
\begin{verbatim}        
expt_list = odesolve(hamiltonian_t, psi0, tlist, 
                     c_op_list, expt_list, H_args)  
\end{verbatim}
\end{footnotesize}
The result of this calculation is shown in Fig.~(\ref{fig:lz_occupation}), which shows the intermediate dynamics of the system in terms of the occupation probabilities of the $\left|0\right>$ and $\left|1\right>$ states. The full code is shown in \ref{sec:lz_code}. Adding the operators \texttt{sigmax, sigmay} and \texttt{sigmaz} to \texttt{expt\_list}, this evolution can also be visualized on the Bloch sphere using QuTiP's built in \texttt{Bloch} class, as demonstrated in Fig.~\ref{fig:lz_bloch}.  The QuTiP code for this figure is given in \ref{sec:bloch_code}.

Although simple, this example illustrates the support for time-dependent Hamiltonians in QuTiP. It is a straightforward exercise to implement an arbitrary time-dependence by changing the definition of the Hamiltonian callback function, or by modifying the time-independent part of the Hamiltonian to correspond to a more complicated quantum system.

\section{Performance}\label{sec:performance}
As with any scientific simulation, the performance of the underlying numerical routines in QuTiP is an important factor to consider when choosing which software to implement in the analysis of the problem at hand.  In simulating quantum dynamics on a classical computer, this is especially important given that creating composite systems using the tensor product leads to an exponential increase in the total Hilbert space dimensionality.  Thus, it is beneficial to compare the performance of QuTiP to the other currently available quantum simulation packages.  In this section we compare the simulation times of the master equation and Monte-Carlo solvers in QuTiP, to the those of the qotoolbox, as a function of Hilbert space size.

\begin{figure}[t]
\begin{center}
\includegraphics[width=8.5cm]{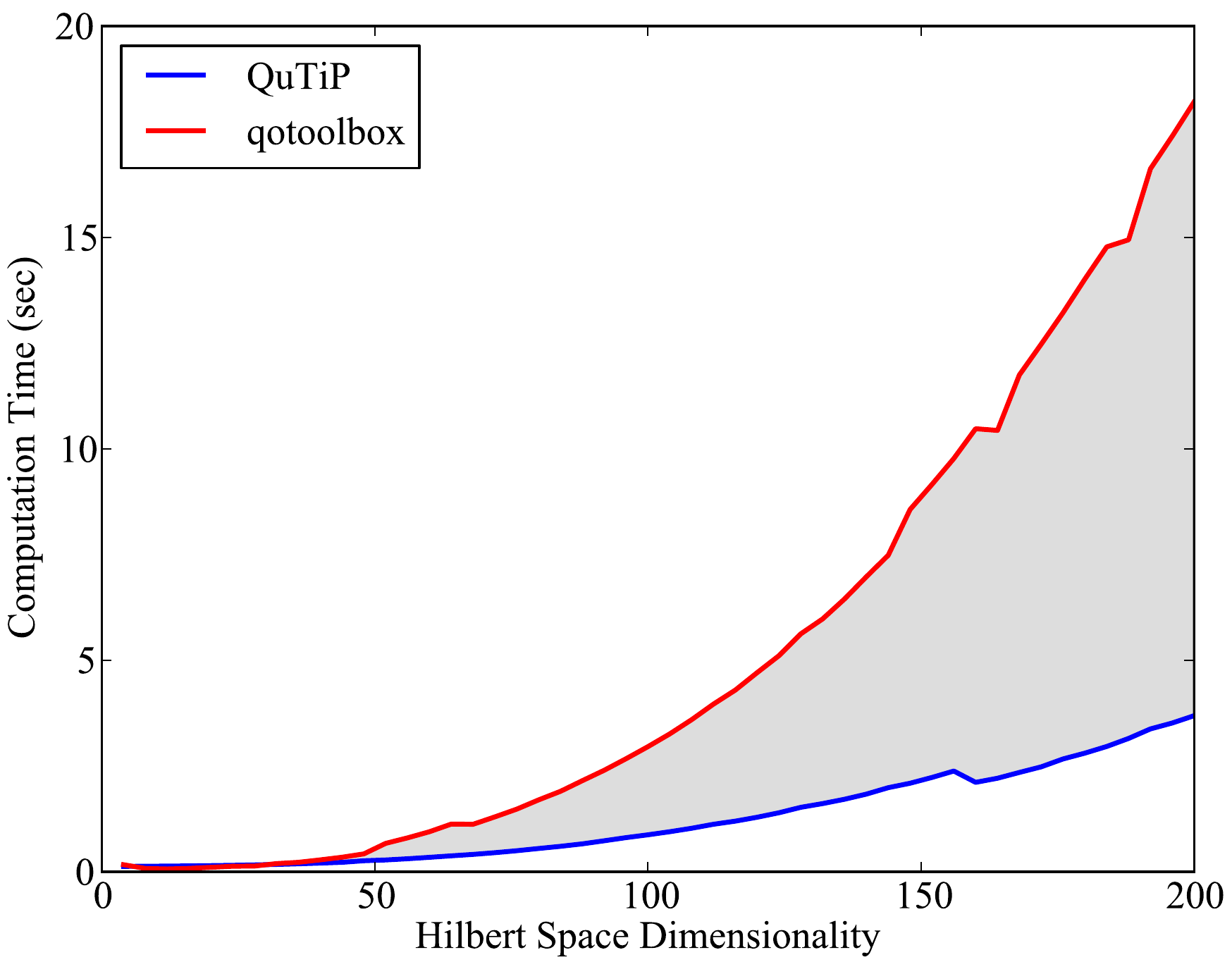}
\caption{(Color) Comparison of computation times, averaged over three trials, for solving Eq.~(\ref{eq:coupled}) in both QuTiP (blue) and the qotoolbox (red) as a function of Hilbert space dimension.  The shaded region highlights the increasing performance benefit from using the QuTiP solver as the dimensionality increases.  Simulations were performed on a quad-core 2.8~Ghz processor.}
\label{fig:odeperform}
\end{center}
\end{figure}

To compare the QuTiP master equation solver, \texttt{odesolve}, to the qotoolbox equivalent (also called \texttt{odesolve}), we evaluate the coupled oscillator equation ($\hbar=1$)
\begin{equation}\label{eq:coupled}
H=\omega_{a}a^{\dag}a+\omega_{b}b^{\dag}b+\omega_{ab}\left(a^{\dag}b+ab^{\dag}\right),
\end{equation}
with $\omega_{a}=\omega_{b}=2\pi$ and $\omega_{ab}=0.1\times2\pi$.  In addition, we consider the situation in which one resonator is damped with a corresponding dissipation rate $g=0.05$.  Here, the initial state of the system is the tensor product of Fock states $|\psi(0)\rangle=|N\rangle_{a}|N-1\rangle_{b}$, where $N$ is the number of states in the truncated Hilbert space for each oscillator. The total time needed to simulate the dynamics over the time range $t\in \left[0,10\right]$ is shown in Fig.~(\ref{fig:odeperform}).  We see that the QuTiP solver easily outperforms the qotoolbox as the Hilbert space dimensionality $D=N^{2}$ increases.  For the largest dimensionality considered in Fig.~(\ref{fig:odeperform}) $D=200$, QuTiP is $\sim 4$ times faster than the qotoolbox single-precision C-code implementation, even though QuTiP is performing double-precision calculations with a small Python overhead.

\begin{figure}[t]
\begin{center}
\includegraphics[width=8.8cm]{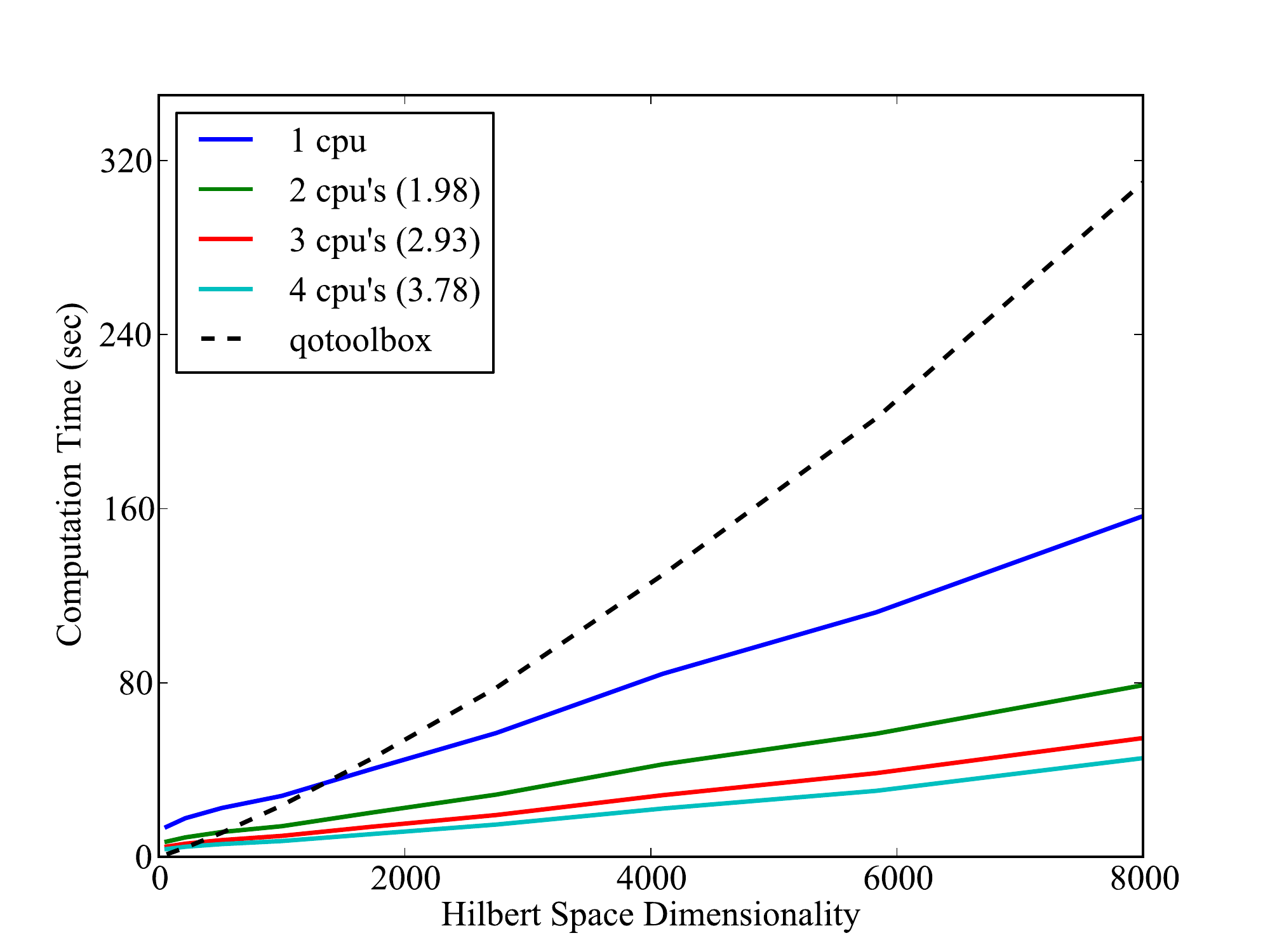}
\caption{(Color) Comparison of computation times, averaged over three runs, between QuTiP and the qotoolbox (dashed) in simulating the trilinear Hamiltonian from Eq.(\ref{eq:trilinear}) for an initial pump mode coherent state with expectation value $\langle N \rangle_{a}=\sqrt[3]{D}/4$ as a function of the Hilbert space dimensionality $D$.  The average performance enhancement from using multiple processors, as compared to the single-processor performance, is given in the legend.  Ideally, this value should be equal to the number of processors. Computations were performed on a quad-core 2.8~Ghz processor.}
\label{fig:mcperform}
\end{center}
\end{figure}

The trilinear Hamiltonian model from Sec.~\ref{sec:tri} provides a useful demonstration of the multiprocessing routines used by the QuTiP \texttt{mcsolve} function.  Here, the independent MC trajectories are run in parallel, with the number of simultaneous trajectories determined by the number of processing cores.  For the large Hilbert spaces associated with the trilinear Hamiltonian, the increased overhead generated from multiprocessing is overcome by the gains in running Monte-Carlo trajectories in parallel.  In Fig.~(\ref{fig:mcperform}), we highlight these performance gains in simulating Eq.~(\ref{fig:trilinear}) by plotting the computation time over a range of Hilbert space dimensions.  For comparison, we also plot the times required for identical simulations using the qotoolbox, which is limited to a single processor.
For this simulation, the qotoolbox outperforms our QuTiP implementation for system sizes $D\lesssim 500$, even when multiple processors are utilized.  This is due to the Python overhead needed to implement the Monte-Carlo algorithm discussed in Sec.~\ref{sec:monte}.  However, as shown in Fig.~(\ref{fig:solver-performance}), using the master equation is more appropriate for systems of this size.  As with the master equation solver, Fig.~(\ref{fig:odeperform}), the benefits of using the QuTiP Monte-Carlo solver become appreciable as the system size increases.  Even for a single-processor, the QuTiP \texttt{mcsolve} routine outperforms the qotoolbox after $D\approx 1500$, where the Python overhead is no longer a limiting factor.  When using multiple processing cores, the Monte-Carlo solver performance gain nearly equals the number of processors available, the ideal situation.

\section{Conclusion}\label{sec:conclusion}

We have presented a new, open-source framework for the numerical simulation of open quantum systems implemented in the Python programming language. This framework is suitable for a wide range of computational problems in quantum systems, including unitary and dissipative time-evolution, spectral and steady-state properties, as well as advanced visualization techniques. In this work we have described the basic structure of the framework, the \texttt{Qobj} class, and the primary evolution solvers, \texttt{odesolve} and \texttt{mcsolve}.  In addition, we have highlighted a number of examples intended to give the reader a flavor of the types of problems for which QuTiP is suitable. For more in-depth documentation, and more elaborate examples using the functions listed in Table~\ref{tbl:list}, we refer the reader to the QuTiP website \cite{qutip}. There, one may download the latest version of this framework, as well as find installation instructions for the most common computer platforms. The version of the framework described in this paper is QuTiP 1.1.3.

As with any initial software release, the performance of QuTiP can likely be improved in the future, with additional portions of the code being optimized with Cython \cite{behnel:2011} or implemented in PyOpenCL \cite{pyopencl}.

\section*{Acknowledgements}
JRJ and PDN were supported by Japanese Society for the Promotion of Science (JSPS) Foreign Postdoctoral Fellowship No.~P11501 and P11202, respectively.  FN acknowledges partial support from the LPS, NSA, ARO, DARPA, AFOSR, National Science Foundation (NSF) grant No. 0726909, Grant-in-Aid for Scientific Research (S), MEXT Kakenhi on Quantum Cybernetics, and the JSPS-FIRST program.

\appendix
\section{QuTiP function list}\label{sec:list}
\onecolumn
\begin{table}[ht]
\begin{footnotesize}
\begin{center}
\begin{tabular}{lll}
\\ \hline\hline
\\
\vspace{0.5mm}
\textit{\small{States}} &
\\
\texttt{basis / fock} &  Creates single basis (Fock) state in Hilbert space.
\\
\texttt{coherent} & Single-mode coherent state with complex amplitude $\alpha$.
\\
\texttt{coherent\_dm} & Coherent state density matrix with complex amplitude $\alpha$.    
\\
\texttt{fock\_dm} & Density matrix representation of a single basis (Fock) state.                                                          
\\
\texttt{qstate} & Tensor product state for any number of qubits in either the ground or excited states.
\\
\texttt{thermal\_dm} & Thermal state density matrix.
\\
\\
\vspace{0.5mm}
\textit{\small{Operators}} &
\\
\texttt{qeye}   & Identity operator.
\\
\texttt{create} & Bosonic creation operator.
\\
\texttt{destroy} & Bosonic annihilation operator.
\\
\texttt{displace}  & Single-mode displacement operator.
\\
\texttt{squeez}  & Single-mode squeezing operator.
\\
\texttt{num}   & Number operator.
\\
\texttt{sigmax}   & Pauli spin-1/2 $\sigma_x$ operator.
\\
\texttt{sigmay}   & Pauli spin-1/2 $\sigma_y$ operator.
\\
\texttt{sigmaz}   & Pauli spin-1/2 $\sigma_z$ operator.
\\
\texttt{sigmap}   & $\sigma_{+}$ operator.
\\
\texttt{sigmam}   & $\sigma_{-}$ operator.
\\
\texttt{jmat}   & Spin-$j$ operator.
\\
\\
\vspace{0.5mm}
\textit{\small{Functions on states}} &
\\
\texttt{entropy\_vn}   & Von-Neumann entropy of a density matrix.
\\
\texttt{expect}   & Calculates the expectation value of an operator.
\\
\texttt{fidelity}   & Calculates the fidelity between two density matrices.
\\
\texttt{ket2dm} & Converts a ket vector to a density matrix.
\\
\texttt{liouvillian}   & Assembles the Liouvillian super-operator from a Hamiltonian and a list of collapse operators.
\\
\texttt{orbital}   & Calculates an angular wave function on a sphere.
\\
\texttt{ptrace}   & Partial trace of composite quantum object.
\\
\texttt{qfunc}   & Husimi-Q function of a given state vector or density matrix.
\\
\texttt{simdiag}   & Simultaneous diagonalization of commuting Hermitian operators.
\\
\texttt{tensor}   & Calculates tensor product from list of input operators or states.
\\
\texttt{tidyup}   & Removes small elements from a quantum object.
\\
\texttt{tracedist}   & Trace distance between two density matrices.
\\
\texttt{wigner}   & Wigner function of a given state vector or density matrix.
\\
\\
\vspace{0.5mm}
\textit{\small{Evolution}} &
\\
\texttt{correlation\_es}   &  Two-time correlation function using exponential series.
\\
\texttt{correlation\_mc}   &  Two-time correlation function using Monte-Carlo method.
\\
\texttt{correlation\_ode}   &  Two-time correlation function using ODE solver.
\\
\texttt{correlation\_ss\_es}   &  Two-time correlation function using quantum regression theorem.
\\
\texttt{essolve}   & State or density matrix evolution using exponential series expansion of ODE.
\\
\texttt{mcsolve}   & Stochastic Monte-Carlo wave function solver.
\\
\texttt{Odeoptions}   & Options class for ODE integrators used by \texttt{mcsolve} and \texttt{odesolve}.
\\
\texttt{odesolve}   & ODE solver for density matrix evolution.
\\
\texttt{propagator}   & Calculates the propagator $U(t)$ for a density matrix or wave function.
\\
\texttt{propagator\_steadystate}   & Steady state for successive applications of the propagator $U(t)$.
\\
\texttt{steadystate}   & Calculates the steady state for the supplied Hamiltonian.
\\
\\
\vspace{0.5mm}
\textit{\small{Utilities}} &
\\
\texttt{about}   & Information on installed version of QuTiP and its dependencies.
\\
\texttt{Bloch}   & Class for plotting vectors and data points on the Bloch sphere.
\\
\texttt{clebsch}   & Calculates a Clebsch-Gordon coefficient.
\\
\texttt{demos}   & Runs built-in demos scripts.
\\
\texttt{eseries}   & Exponential series representation of a time-dependent quantum object.
\\
\texttt{ode2es}   & Exponential series describing the time evolution of an initial state.
\\
\texttt{parfor}   & Parallel execution of a for-loop over a single-variable.
\\
\texttt{Qobj}   & Class for creating user-defined quantum objects.
\\
\texttt{sphereplot}   & Plots an array of values on a sphere.
\\
\\ 
\hline\hline

\end{tabular}
\end{center}
\caption{List of user-accessible functions in QuTiP.  Additional information about each function may be obtained by calling `\textit{function\_name}?' from the Python command-line, or by going to the QuTiP website \cite{qutip}.}
\label{tbl:list}
\end{footnotesize}
\end{table}
\twocolumn

\section{QuTiP codes}\label{ap:codes}
In this section we display the QuTiP codes underlying the calculations performed in generating Figures~\ref{fig:nonrwa}, \ref{fig:decay}, \ref{fig:iswap}, \ref{fig:jc}, \ref{fig:trilinear}, \ref{fig:lz_occupation}, and \ref{fig:lz_bloch}.  For brevity, the code segments associated with plotting have been omitted.  The codes, in their entirety, may be viewed at the QuTiP website \cite{qutip}.

\subsection{Figure~\ref{fig:nonrwa}: Non-RWA Jaynes-Cummings Model}\label{sec:nonrwa_code}
\begin{footnotesize}
\begin{verbatim}
from qutip import *
## set up the calculation ## 
wc = 1.0 * 2 * pi # cavity frequency
wa = 1.0 * 2 * pi # atom frequency
N = 20            # number of cavity states
g = linspace(0, 2.5, 50)*2*pi # coupling strength vector
## create operators ## 
a  = tensor(destroy(N), qeye(2))
sm = tensor(qeye(N), destroy(2))
nc = a.dag() * a
na = sm.dag() * sm
## initialize output arrays ##
na_expt = zeros(len(g))
nc_expt = zeros(len(g))
## run calculation ## 
for k in range(len(g)):
    ## recalculate the hamiltonian for each value of g ## 
    H = wc*nc+wa*na+g[k]*(a.dag()+a)*(sm+sm.dag())
    ## find the groundstate ## 
    ekets, evals = H.eigenstates()
    psi_gnd = ekets[0]
    ## expectation values ## 
    na_expt[k] = expect(na, psi_gnd) # qubit occupation
    nc_expt[k] = expect(nc, psi_gnd) # cavity occupation 
## Calculate Wigner function for coupling g=2.5 ## 
rho_cavity = ptrace(psi_gnd,0) # trace out qubit
xvec = linspace(-7.5,7.5,200)
## Wigner function ## 
W = wigner(rho_cavity, xvec, xvec)
\end{verbatim}
\end{footnotesize}

\subsection{Figure~\ref{fig:decay}: Monte-Carlo relaxation in a thermal environment}\label{sec:thermal_code}
\begin{footnotesize}
\begin{verbatim}
from qutip import *
N=5             # number of basis states to consider
a=destroy(N)    # cavity destruction operator
H=a.dag()*a     # harmonic oscillator Hamiltonian
psi0=basis(N,1) # initial Fock state with one photon
kappa=1.0/0.129 # coupling to heat bath
nth= 0.063      # temperature with <n>=0.063
## collapse operators ## 
c_op_list = []
## decay operator ## 
c_op_list.append(sqrt(kappa * (1 + nth)) * a)
## excitation operator ## 
c_op_list.append(sqrt(kappa * nth) * a.dag())
## run simulation ## 
ntraj=904 # number of MC trajectories
tlist=linspace(0,0.6,100)
mc = mcsolve(H,psi0,tlist,ntraj,c_op_list, [])
me = odesolve(H,psi0,tlist,c_op_list, [a.dag()*a])
## expectation values ## 
ex1=expect(num(N),mc[0])
ex5=sum([expect(num(N),mc[k]) for k in range(5)],0)/5
ex15=sum([expect(num(N),mc[k]) for k in range(15)],0)/15
ex904=sum([expect(num(N),mc[k]) for k in range(904)],0)/904
\end{verbatim}
\end{footnotesize}

\subsection{Figure~\ref{fig:iswap}: Dissipative $i$-SWAP gate}\label{sec:iswap_code}

\begin{footnotesize}
\begin{verbatim}
from qutip import *
g  = 1.0 * 2 * pi   # coupling strength
g1 = 0.75           # relaxation rate
g2 = 0.05           # dephasing rate
n_th = 0.75         # bath temperature
T = pi/(4*g) 
H = g * (tensor(sigmax(), sigmax()) +
         tensor(sigmay(), sigmay()))
psi0 = tensor(basis(2,1), basis(2,0))     
c_op_list = []
## qubit 1 collapse operators ## 
sm1 = tensor(sigmam(), qeye(2))
sz1 = tensor(sigmaz(), qeye(2))
c_op_list.append(sqrt(g1 * (1+n_th)) * sm1)
c_op_list.append(sqrt(g1 * n_th) * sm1.dag())
c_op_list.append(sqrt(g2) * sz1)
## qubit 2 collapse operators ## 
sm2 = tensor(qeye(2), sigmam())
sz2 = tensor(qeye(2), sigmaz())
c_op_list.append(sqrt(g1 * (1+n_th)) * sm2)
c_op_list.append(sqrt(g1 * n_th) * sm2.dag())
c_op_list.append(sqrt(g2) * sz2)
## evolve the system ## 
tlist = linspace(0, T, 100)
rho_list  = odesolve(H, psi0, tlist, c_op_list, [])
rho_final = rho_list[-1]
## calculate expectation values ## 
n1 = expect(sm1.dag() * sm1, rho_list)
n2 = expect(sm2.dag() * sm2, rho_list)     
## calculate the fidelity ## 
U = (-1j * H * pi / (4*g)).expm()
psi_ideal = U * psi0
rho_ideal = psi_ideal * psi_ideal.dag()
f = fidelity(rho_ideal, rho_final) 
\end{verbatim}
\end{footnotesize}

\subsection{Figure~\ref{fig:jc}: Dissipative Jaynes-Cumming model}\label{sec:jc_code}

\begin{footnotesize}
\begin{verbatim}
from qutip import *
N = 5                      # number of cavity states
omega0 = epsilon = 2 * pi  # frequencies
g = 0.05 * 2 * pi          # coupling strength
kappa = 0.005              # cavity relaxation rate
gamma = 0.05               # atom relaxation rate
n_th = 0.75                # bath temperature 
## Hamiltonian and initial state ## 
a  = tensor(destroy(N), qeye(2))
sm = tensor(qeye(N), destroy(2))
sz = tensor(qeye(N), sigmaz())
H  = omega0 * a.dag() * a + 0.5 * epsilon * sz 
     + g * (a.dag() * sm + a * sm.dag())
psi0 = tensor(fock(N,0), fock(2,1)) # excited atom
## Collapse operators ## 
c_ops = []
c_ops.append(sqrt(kappa * (1+n_th)) * a)
c_ops.append(sqrt(kappa * n_th) * a.dag())
c_ops.append(sqrt(gamma) * sm)
## Operator list for expectation values ## 
expt_ops = [a.dag() * a, sm.dag() * sm]
## Evolution of the system ## 
tlist = linspace(0, 10, 100)
expt_data = odesolve(H, psi0, tlist, c_ops, expt_ops)
\end{verbatim}
\end{footnotesize}

\subsection{Figure~\ref{fig:trilinear}: Trilinear Hamiltonian}\label{sec:trilinear_code}
\begin{footnotesize}
\begin{verbatim}
from qutip import *
N=17 # number of states for each mode
## damping rates ## 
g0=g2=0.1
g1=0.4
alpha=sqrt(10) # initial coherent state alpha
tlist=linspace(0,4,201) # list of times
ntraj=1000#number of trajectories
## lowering operators ## 
a0=tensor(destroy(N),qeye(N),qeye(N))
a1=tensor(qeye(N),destroy(N),qeye(N))
a2=tensor(qeye(N),qeye(N),destroy(N))
## number operators ## 
n0,n1,n2=[a0.dag()*a0,a1.dag()*a1,a2.dag()*a2]
## dissipative operators ## 
C0,C1,C2=[sqrt(2.0*g0)*a0,sqrt(2.0*g1)*a1,sqrt(2.0*g2)*a2]
## initial state ## 
psi0=tensor(coherent(N,alpha),basis(N,0),basis(N,0))
## trilinear Hamiltonian ## 
H=1j*(a0*a1.dag()*a2.dag()-a0.dag()*a1*a2)
## run Monte-Carlo ## 
avgs=mcsolve(H,psi0,tlist,ntraj,[C0,C1,C2],[n0,n1,n2])
## run Schrodinger ## 
reals=mcsolve(H,psi0,tlist,1,[],[n0,n1,n2])
\end{verbatim}
\end{footnotesize}

\subsection{Figure~\ref{fig:lz_occupation}: Landau-Zener transitions}\label{sec:lz_code}

\begin{footnotesize}
\begin{verbatim}
from qutip import *
## callback function for time-dependence ## 
def hamiltonian_t(t, args):
    H0 = args[0]
    H1 = args[1]
    return H0 + t * H1
delta = 0.5 * 2 * pi 
v = 2.0 * 2 * pi # sweep rate
## arguments for Hamiltonian ## 
H0 = delta/2.0 * sigmax()
H1 = v/2.0 * sigmaz()
H_args = (H0, H1)
psi0 = basis(2,0)
## expectation operators ## 
sm = destroy(2)
sx=sigmax();sy=sigmay();sz=sigmaz()
expt_op_list = [sm.dag() * sm,sx,sy,sz]
## evolve the system ## 
tlist = linspace(-10.0, 10.0, 1500)
expt_list = odesolve(hamiltonian_t, psi0, tlist, 
                     [], expt_op_list, H_args)  
\end{verbatim}
\end{footnotesize}

\subsection{Figure~\ref{fig:lz_bloch}: Bloch sphere representation of Landau-Zener transition}\label{sec:bloch_code}
Following the code from \ref{sec:lz_code}:
\begin{footnotesize}
\begin{verbatim}
import matplotlib as mpl
from matplotlib import cm
## create Bloch sphere instance ## 
b=Bloch()
## normalize colors to times in tlist ## 
nrm=mpl.colors.Normalize(-2,10)
colors=cm.jet(nrm(tlist))
## add data points from expectation values ## 
b.add_points([p_ex[1],p_ex[2],-p_ex[3]],'m')
## customize sphere properties ## 
b.point_color=list(colors)
b.point_marker=['o']
b.point_size=[8]
b.view=[-9,11]
b.zlpos=[1.1,-1.2]
b.zlabel=['$\left|0\\right>_{f}$','$\left|1\\right>_{f}$']
## plot sphere ## 
b.show()
\end{verbatim}
\end{footnotesize}

\bibliographystyle{model1a-num-names}
\bibliography{text}

\end{document}